\documentclass[twocolumn]{aa}  

\usepackage{graphicx}
\usepackage{multicol}
\usepackage{amssymb}
\usepackage{dsfont}
\usepackage{makecell}
\usepackage{subcaption}
\usepackage{txfonts}
\usepackage{multirow}
\begin{document} 

    \title{Gravitational collapse at low to moderate Mach numbers: 
    \\The relationship between star formation efficiency and the fraction of mass in the massive object}

%   \title{The formation of massive objects via gravitational collapse at low to moderate Mach numbers: relation between the star formation efficiency and the fraction of mass in the massive object}

   \titlerunning{Formation of massive objects}

%   \subtitle{I. Overviewing the $\kappa$-mechanism}

   \author{Jorge Saavedra-Bastidas
          \inst{1}
          \and
          Dominik R.G. Schleicher\inst{1}
\and
Ralf S. Klessen\inst{2,3}
\and
Sunmyon Chon\inst{4,5} 
\and
Kazuyuki Omukai\inst{5}
\and
Thomas Peters\inst{4}
\and 
Lewis R. Prole\inst{6}
\and
Basti\'an Reinoso\inst{7}
\and
Rafeel Riaz\inst{8}
\and 
Paulo Solar\inst{9}
}

   \institute{Departamento de Astronom\'ia, Facultad Ciencias F\'isicas y Matem\'aticas,    Universidad de Concepci\'on, Av. Esteban Iturra s/n Barrio Universitario, Casilla 160-C, Concepci\'on, Chile
             \and 
             Universit\"at Heidelberg, Zentrum f\"ur Astronomie, Institut f\"ur theoretische Astrophysik, Albert-Ueberle Str. 2, 69120 Heidelberg, Germany
             \and 
             Universit\"at Heidelberg, Interdisziplin\"ares Zentrum f\"ur wissenschaftliches Rechnen, Im Neuenheimer Feld 205, 69120 Heidelberg, Germany
             \and
             Max-Planck-Institut f\"ur Astrophysik, Karl-Schwarzschild-Str. 1, D-85748 Garching, Germany
            \and 
             Astronomical Institute, Graduate School of Science, Tohoku University, Aoba, Sendai 980-8578, Japan
            \and
             Centre for Astrophysics and Space Sciences Maynooth, Department of Theoretical Physics, Maynooth University, Maynooth, Ireland
             \and
             Universit\"at Heidelberg, Zentrum für Astronomie, Institut f\"ur Theoretische Astrophysik, Albert-Ueberle-Str. 2, 69120 Heidelberg, Germany
             \and
             Centro de investigaci\'on en Astronom\'ia, Facultad de Ingeniería, Ciencia y Tecnología, Universidad Bernardo O'Higgins, Av. Viel 1497, Santiago, Chile \\
             \and
             Hamburger Sternwarte, Universit\"at Hamburg, Gojenbergsweg 112, 21029 Hamburg, Germany
             }

   \date{Received September 15, 1996; accepted March 16, 1997}

% \abstract{}{}{}{}{} 
% 5 {} token are mandatory
 
  \abstract
   {The formation of massive objects via gravitational collapse is relevant both for explaining the origin of the first supermassive black holes and in the context of massive star formation. Here, we analyze simulations of the formation of massive objects pursued by different groups and in various environments, concerning the formation of supermassive black holes, primordial stars, as well as present-day massive stars. {We focus here particularly on the regime of small virial parameters, i.e., low ratios of the initial kinetic to gravitational energy, low to moderate Mach numbers, and the phase before feedback is very efficient.} We compare the outcomes of collapse under different conditions using dimensionless parameters, particularly the star formation efficiency $\epsilon_*$, the fraction $f_*$ of mass in the most massive object relative to the total stellar mass, and the fraction $f_{\rm tot}$ of mass of the most massive object as a function of the total mass. We find that in all simulations analyzed here, $f_{\rm tot}$ increases as a function of $\epsilon_*$, although the steepness of the increase depends on the environment. The relation between $f_*$ and $\epsilon_*$ is found to be more complex and also strongly depends on the number of protostars present at the beginning of the simulations. We show that a collision parameter, estimated as the ratio of the system size divided by the typical collision length, allows us to approximately characterize whether collisions will be important. A high collision parameter implies a steeper increase in the relation between $f_{\rm tot}$ and $\epsilon_*$. We analyze the statistical correlation between the dimensionless quantities using the Spearman coefficient and further confirm via a machine learning analysis that good predictions of $f_*$ can be obtained from $\epsilon_*$ together with a rough estimate of the collision parameter. This suggests that a good estimate of the mass of the most massive object can be obtained once the maximum efficiency for a given environment is known and an estimate for the collision parameter has been determined.}
   

   \maketitle
%
%-------------------------------------------------------------------

%%%% S E C T I O N %% I %%%%%%%%%%%%%%%%%%%%%%%%%%%%%%%%%%%%%%%%%%%%%%%%%%%%%%%%%%%%%%%%%%%%%%%%%%%%%%%%%%%%%%%%%%%%%%%%%%%%%%%%%%%%%%%%%%%%%%%%%%%%%%%%%%%%%%
\section{Introduction}
% 1-About SMBHs and their formation scenarios ---------------------------------------------------------------------------------------------------------------%      
The formation of massive objects via gravitational collapse is an important subject of investigation to explain the origin of supermassive black holes and in the context of star formation in general. In the early Universe, more than 200 supermassive black holes are currently known at $z\gtrsim6$ \citep{Banados2016, Inayoshi2020}. Several different origins have been proposed for the formation of their seeds, typically considering different variations of the pathways proposed by \citet{Rees1984}. All of these pathways involve gravitational collapse and gravitational instability, and understanding the outcome of gravitational collapse is thus very likely at the heart of resolving this problem. One of the proposed pathways is the so-called direct collapse, leading to the formation of a single massive object from a massive cloud \citep[e.g.,][]{Bromm2003, Koushiappas2004, Begelman06, Wise2008a, Schleicher10b, Latif2013BH, Inayoshi2014, Regan2017, Woods2019}. Recent observations with the James Webb Space Telescope (JWST) indeed reveal an object with a $\sim10^6$~M$_\odot$ that may have formed either due to direct collapse or from smaller seeds experiencing significant super-Eddington accretion \citep{Larson2023}. We note here, however, that even under strongly idealized conditions, at least some fragmentation was still found, suggesting that the formation of the central massive object at least must be accompanied by the formation of additional clumps \citep[e.g.,][]{Latif2013BH, Suazo2019, Latif2021, Prole2024}.

% 2-About fragmentation -------------------------------------------------------------------------------------------------------------------------------------%      
Fragmentation in the context of the direct collapse scenario has meanwhile been studied under a wider range of conditions and was found to be difficult to avoid. The presence of H$_2$ cooling, for example, was shown to decrease the mass of the most massive object and favor fragmentation \citep{Latif2014mol}. The presence of small amounts of dust grains can trigger fragmentation at densities above $10^{10}$~cm$^{-3}$ and lead to the formation of rather low-mass clumps \citep{Omukai2008, Latif2016}. On the other hand, simulations also have shown that fragmentation does not necessarily impede the formation of a very massive object, as the clumps may also merge again \citep[e.g.,][]{Suazo2019, Grete2019, Prole2024}. This behavior can be understood as a result of clump migration in self-gravitating accretion disks \citep{LatifSchleicher2015}.

% 3-Simulations including fragmentation and mergers ---------------------------------------------------------------------------------------------------------%     
As a result of these investigations, it has become more common to consider models that allow both fragmentation and mergers to occur, considering growth via accretion as well as through mergers of fragments. A toy model for the dynamical evolution of a primordial protostellar cluster embedded in gas has been presented by \citet{Boekholt2018}, showing that it could lead to the formation of massive objects of $10^5$~M$_\odot$. \citet{Tagawa2020} presented similar results employing semi-analytic calculations. \citet{Chon2020} presented numerical simulations starting from cosmological initial conditions, showing that a central massive object can still accrete efficiently even in the presence of fragmentation. An analytic assessment of the formation of a massive object via collisions and accretion, including the effect of feedback, has been provided by \citet{Schleicher2022}. \citet{Reinoso2023} presented numerical simulations following the dynamics of gas and protostars for more than $100,000$~years, showing that objects of $\sim10^4$~M$_\odot$ can form. 

Similar simulations including the formation of fragments and subsequent mergers have been recently pursued e.g. by \citet{Prole2022, Prole2023} in the context of Population III (Pop.~III) star formation \citep[for a review, see][]{Klessen2023}. \citet{Riaz2020}, on the other hand, employed a parameterized equation of state to explore how variations in the equation of state, in particular the initial temperature, affect the initial mass function, including the formation of massive stars via fragmentation and accretion. Mass segregation can further play an additional role in changing the distribution of stars within the cluster, with the heavier ones dynamically sinking towards the center of the distribution \citep[e.g.,][]{Bonnell1998, Allison2009, Olczak2011}.

The relevance of many of these phenomena is also known from the context of massive star formation \citep[e.g.,][]{Bonnell2004, Peters2010, Seifried2011, Myers2013}. Investigating the relation between envelope masses and final stellar mass, \citet{Bonnell2004} found no correlation between the clump masses and the final masses. However, the masses of the less-than-solar-mass stars were closely related to the envelope masses at the time of protostar formation. In the case of stars with more than one solar mass, on the other hand, a significant fraction of their final mass was obtained from accretion outside their original envelope. 

% 4-Alternative formation paths------------------------------------------------------------------------------------------------------------------------------%      
Other formation channels of massive objects, particularly black holes, are based on collisions only. These could occur in dense star clusters at low metallicities \citep{Omukai2008, Devecchi2009, Katz2015, Sakurai2017,  Reinoso2018, Reinoso2020, Vergara2021} or  Nuclear Star Clusters \citep[NSCs, e.g.,][]{Escala2021, Vergara2023}. A relevant source could be even the mergers in black hole clusters, as initially investigated by \citet{Quinlan1987, Quinlan1990}. In those simulations, it was found that the black hole clusters would need to be extremely dense so that mergers via gravitational wave emission dominate over the ejection through three-body interactions. However, \citet{Davies2011} have shown that sufficient steepening of the black hole distribution may occur in the case that a gas inflow into the center acts as an external potential, thereby decreasing the timescale for gravitational wave emission. \citet{Lupi2014} have found the black hole population resulting from this channel to be relevant, and similarly, \citet{Kroupa2020} have shown that the latter provides an efficient mechanism to explain the observed supermassive black holes. 

% 5-Our scientific aim --------------------------------------------------------------------------------------------------------------------------------------%      
In this paper, our interest concerns particularly the formation of massive central objects from gas, including the processes of fragmentation and subsequent mergers. {We focus here particularly on situations with small virial parameters $\alpha$, i.e. low ratios of initial kinetic turbulent to gravitational energy and only low to moderate ($M\sim1$) Mach numbers. The focus of this investigation  further concerns the phase before feedback is very efficient. The relations derived here can thus be expected to hold as long as these assumptions are being fulfilled; thereby allowing to obtain at least an upper limit for the mass of the most massive object that is obtained before strong feedback sets in. In particular, we}  analyze how much the formation of massive objects differs between different simulations, including different types of environments. For this purpose, we consider the mass of the most massive object normalized both via the total stellar mass in the cluster as well as the total mass in the cluster, considering both gas and stars, as a function of the star formation efficiency that we defined as the stellar mass divided by the initial gas mass, thus describing how much of the initial gas mass has already been converted into stars. {The suite of simulations employed in this work is mostly employing piecewise-polytropic equations of state and not explicitly modeling the cooling. We may therefore miss some effects that could occur in the limit when the cooling time is comparable to or larger than the free-fall time of the simulations.}

% 6-Paper index ---------------------------------------------------------------------------------------------------------------------------------------------%      
The data we consider here is summarized in Section~\ref{data}. A visual comparison and analysis of the results is given in Section~\ref{comparison}. In Section~\ref{ML}, we present a machine learning analysis to determine the correlation within the quantities that we analyzed here. A final summary and discussion is given in Section~\ref{discussion}.

%%%% S E C T I O N %% II %%%%%%%%%%%%%%%%%%%%%%%%%%%%%%%%%%%%%%%%%%%%%%%%%%%%%%%%%%%%%%%%%%%%%%%%%%%%%%%%%%%%%%%%%%%%%%%%%%%%%%%%%%%%%%%%%%%%%%%%%%%%%%%%%%%%%
% Table 1: Simulations summary ------------------------------------------------------------------------------------------------------------------------------%      
\begin{table*}[!t]
    \centering
    \begin{tabular}{lclc}
    \hline 
     {\bf Reference}     & {\bf Type of simulation}                                       & {\bf Explored parameter}       & {\bf N$_{sim}$} \\ \hline
     \citet{Reinoso2023} & embedded primordial protocluster, $10^4-3\times10^4$~M$_\odot$ & \makecell[l]{gas mass, 
                                                                                                       \\stellar physics}  & 24  \\ 
     Solar et al. (2024-in progress).   & embedded primordial protocluster, $3\times10^4$~M$_\odot$      & gas temperature                & 10   \\
     \citet{Chon2020}    & massive gas cloud, $2\times10^4$~M$_\odot$                     & metallicity                    & 5   \\
     \citet{Prole2022b}  & primordial collapse, $\sim1800$~M$_\odot$                      & magnetic field                 & 6   \\
     \citet{Prole2023}   & primordial star formation in dark matter halos                 & dark matter halo               & 5   \\
     \citet{Peters2010}  & massive star formation, $1000$~M$_\odot$ cloud                 & -                              & 1   \\
     \citet{Riaz2020}    & star formation, $30$~M$_\odot$ cloud                           & initial temperature            & 5   \\ 
     \hline
    \end{tabular}
    \caption{Summary of the simulations employed here.}\label{tabsims}
\end{table*}
%------------------------------------------------------------------------------------------------------------------------------------------------------------%
\section{Simulation data}\label{data}
% 1-Introduction --------------------------------------------------------------------------------------------------------------------------------------------%      
In the following, we describe the different simulation data that we employ here in this comparison. A summary of these is also provided in Table~\ref{tabsims}.
% 2-B.Reinoso files -----------------------------------------------------------------------------------------------------------------------------------------%
The simulations of \citet{Reinoso2023} have been performed with the AMUSE framework\footnote{AMUSE: https://www.amusecode.org/} \citep{Pelupessy2013}, modeling embedded primordial protostellar clusters employing a simplified equation of state, where hydrodynamics is treated using the smoothed particle hydrodynamics (SPH) code FI \citep{Hernquist1989}, the stellar dynamics with the N-body code PH4 \citep{McMillan1996}, which are coupled using the BRIDGE method \citep{Fujii2007}. Protostars are treated as sink particles and modeled following \citet{Hubber2013}. Both gas and the protostellar distribution follow a Plummer sphere \citep{Plummer1911} with a Plummer radius of $0.077$~pc. They explore total gas masses of $10^4$~M$_\odot$ and $3\times10^4$~M$_\odot$, employing $2^{18}$ SPH particles. The initial number of protostars is 256, with each of them having an initial mass of $0.1$~M$_\odot$. The simulations include an initial spectrum of non-compressive Kolmogorov-type turbulence corresponding to a Mach number of $\mathcal{M}=1$, {and the virial parameter $\alpha$ ranges between $0.33$ and $1$ depending on the mass of the cluster.} For the mass-radius relation of the protostars, a fit to the relations calculated by \citet{Hosokawa2009, Hosokawa2012, Hosokawa2013} is being employed. Supermassive protostars are known to contract when their average accretion rate falls below a value of $\sim0.04$~M$_\odot$~yr$^{-1}$ \citep{Hosokawa2013, Schleicher2013, Haemmerle2018}. In the simulations of \citet{Reinoso2023}, this is assumed to happen when the accretion rate falls below the critical value of $X$ Kelvin-Helmholtz timescales, where they explore $X=10$ and $X=100$. The simulations have been evolved for $200,000$~years. For each parameter combination, they perform six simulations varying the initial random seed for the turbulence. For clouds of  $10^4$~M$_\odot$, they show the formation of massive objects in the range of $\sim2000-5000$~M$_\odot$. For clouds with $3\times10^4$~M$_\odot$, the masses of the most massive object are in the range of $\sim21000-27000$~M$_\odot$. The accretion radius of the sink particles is calculated iteratively via the interaction zone as defined by \citet{Hubber2013}, with a lower limit of $10$~AU and an upper limit of $500$~AU.

% 3-P.Solar files -------------------------------------------------------------------------------------------------------------------------------------------%
The simulations by \citet{Solar2022} was the first explorations employing the same methodology as \citet{Reinoso2023} but focus on clouds with $3\times10^3$~M$\odot$. Now in Solar et al. (2024-in progress), they pursue a systematic exploration of the effect of the initial gas temperature on the evolution of the cloud with a mass of $3\times10^4$~M$\odot$, exploring initial temperatures of $500$~K, $1000$~K, $3000$~K, $5000$~K, and $8000$~K. While the equation of state is initially assumed to be isothermal, it becomes adiabatic for densities larger than $10^{15}$~cm$^{-3}$. They find the mass of the most massive object to increase with decreasing temperature, due to the increasing instability of the cloud, with masses in the range of $\sim19000-28000$~M$_\odot$ after $30.000$~years. {The initial turbulent Mach number in the simulations is equal to $1$, and the virial parameter ranges between $0.03$ and $0.33$.}

% 4-S.Chon files --------------------------------------------------------------------------------------------------------------------------------------------%
\citet{Chon2020} pursued a suite of hydrodynamical simulations employing the SPH code Gadget-2\footnote{Gadget-2: https://wwwmpa.mpa-garching.mpg.de/galform/gadget/} \citep{Springel05}. They extract a massive compact gas cloud from the cosmological simulation of \citet{Chon2016, Chon2018}, containing gas particles inside a radius of $7\times10^6$~AU from the center of the cloud with a total gas mass of $2\times10^4$~M$_\odot$. The equation of state is precalculated following the one-zone models of \citet{Omukai2008} for the chemical evolution, exploring metallicities of $Z/Z_\odot=10^{-3}$, $10^{-4}$, $10^{-5}$, $5\times10^{-6}$ and $10^{-6}$. Sink particles are formed when the density reaches $2\times10^{16}$~cm$^{-3}$. The simulations have been evolved for $\sim10.000$~years. For the mass of the most massive object, they obtained $\sim300$~M$_\odot$ in case of $Z/Z_\odot=10^{-3}$ where the gas is affected by metal line cooling, while it reaches $\sim8000$~M$_\odot$ in simulations with lower metallicities. The adopted accretion radius of the sink particles is equal to the protostellar radius. {The cloud shows transsonic turbulence with a virial parameter $\alpha\leq1$.}

% 5-L.Prole 2022 files --------------------------------------------------------------------------------------------------------------------------------------%
\citet{Prole2022b}  have performed two-stage zoom-in magneto-hydrodynamics (MHD) simulations of primordial star formation with the moving mesh code Arepo\footnote{Arepo code: https://arepo-code.org/wp-content/userguide/index.html} \citep{Springel2010}. The simulations start from Bonnor-Ebert spheres \citep{Ebert1955,Bonnor1956} with a central density of $2\times10^{-20}$~g~cm$^{-3}$ and a radius $R_{BE}=1.87$~pc. The initial condition included a turbulent velocity field with a turbulent power spectrum $\propto k^{-2}$, which is normalized to obtain a ratio of turbulent to gravitational energy of $0.05$. {The turbulence is subsonic and the turbulent Mach number of order $0.3$.} The chemistry and thermodynamics have been solved employing an updated version of the chemical model from \citet{Clark2011}. The authors present both purely hydrodynamical simulations, as well as simulations including tangled magnetic fields, following a Kazantsev spectrum of $k^{3/2}$ \citep{Kazantsev68}, assuming a saturated magnetic field in equipartition with the turbulent velocity. The MHD equations are solved following \citet{Pakmor2011}. The authors have explored critical densities for sink particle formation of $10^{-8}$~g~cm$^{-3}$, $10^{-9}$~g~cm$^{-3}$ and $10^{-10}$~g~cm$^{-3}$, After reaching the critical density for sink formation, the simulations are evolved for about $\sim1500$~years. Typically, 10-20 fragments form in each simulation, with the total protostellar masses varying between $40-80$~M$_\odot$. The most massive protostars in the simulations reach about $\sim10$~M$_\odot$. The sink accretion radius was adopted as the Jeans length at the sink creation density. 

% 6-L.Prole 2023 files --------------------------------------------------------------------------------------------------------------------------------------%
\citet{Prole2023} presented cosmological simulations with the Arepo code \citep{Springel2010} including hydrodynamics and primordial chemistry, modeling gravitational collapse in dark matter halos with masses of $3-9\times10^6$~M$_\odot$ within a cosmological box of $1$~Mpc$/h$, were $h=0.6774$ is the reduced Hubble constant. The chemical network was based on an updated version of the appendix from \citet{Clark2011}. They further considered the photo-dissociation of molecular hydrogen through Lyman-Werner (LW) photons with $11.2-13.6$~eV can dissociate H$_2$ via the two-step Solomon process \citep{Field1966, Stecher67}. They explore Lyman-Werner backgrounds of $J_{21}=0$, $0.1$ and $0.01$, where $J_{21}=1$ implies a Lyman-Werner band intensity of $10^{-21}$~erg~s$^{-1}$~cm$^{-2}$~Hz$^{-1}$~sr$^{-1}$. {From the cosmological initial conditions, turbulence formed dynamically during the gravitational collapse, producing typical turbulent Mach numbers of order $0.3$ and typical virial parameters of order $0.05$. } They adopt a critical density for sink particle formation of $10^{18}$~cm$^{-3}$. The simulations have been evolved for $300$~years after the formation of the first sink particle. The number of sink particles formed in the different halos varies between $5$ and $30$. The total mass in sinks reaches $10-30$~M$_\odot$, and the mass of the most massive sink is comparable, suggesting that few sink particles contain a significant fraction of the total stellar mass. The sink accretion radius was taken to be the protostellar radius estimated through recipes of \citet{Stahler1986}.

% 7-T.Peters file -------------------------------------------------------------------------------------------------------------------------------------------%
\citet{Peters2010b, Peters2010} presented 3D radiation-hydrodynamical simulations of massive star formation including heating by ionizing and non-ionizing radiation using the adaptive-mesh code FLASH\footnote{FLASH code: https://flash.rochester.edu/site/flashcode/} \citep{fryxell00}. The simulation starts from a $1000$~M$_\odot$ molecular cloud with a core density of $1.27\times10^{-20}$~g~cm$^{-3}$ within a radius of $0.5$~pc, falling of as $r^{-3/2}$ until $1.6$~pc. The temperature of the cloud is $T = 30$~K. It is assumed to be in solid body rotation with an angular velocity $\omega=1.5\times10^{-14}$~s$^{-1}$, corresponding to a ratio of rotational to gravitational energy of $0.05$. {Turbulence was not explicitly included in the initial condition.} Sink particles are created at a density of $7\times10^{-16}$~g~cm$^{-3}$. We focus here on their run B, which includes multiple sinks and radiative feedback. In total, it leads to the formation of $25$ sink particles, a total stellar mass of $125.56$~M$_\odot$, where the most massive star reaches $23.39$~M$_\odot$ after $0.74$~Myr.

% 8-R.Riaz --------------------------------------------------------------------------------------------------------------------------------------------------%
Finally, we consider the simulations by \citet{Riaz2020} exploring how fragmentation and accretion affect the stellar initial mass function for varying temperatures of the gas, employing the SPH code GRADSPH\footnote{GRADSPH code: https://zbmath.org/software/1046} \citep{Vanaverbeke2009}. They consider the gravitational collapse of a gas cloud with $30$~M$_\odot$, a radius of $0.168$~pc and a cloud density of $10^{-19}$~g~cm$^{-3}$. They assume rigid rotation with an angular velocity of $2.912\times10^{-14}$~s$^{-1}$, corresponding to a ratio of rotational to gravitational energy of $1\%$. The initial temperature is varied between $10$ and $50$~K, corresponding to ratios of the thermal to gravitational energy between $0.115$ and $0.578$. {The simulations included initial turbulence with a Mach number of $1$, implying virial parameters in the range of $0.115-0.578$.} They employed a barotropic equation of state that was initially isothermal but became adiabatic at the density at which the gas is expected to become optically thick to dust emission following \citet{Omukai05}. Sink particle formation is modeled following \citet{Hubber2013}, employing a constant sink radius of $1$~AU that is always larger than the Jeans length. The simulations are evolved until $15\%$ of the gas mass is converted into stellar mass. The number of sink particles at the end of the simulation varies between $7$ and $30$, and the mean mass per sink particle is between $0.15$ and $0.64$~M$_\odot$. The mass of the most massive sink varies between $0.7$ and $1.3$~M$_\odot$. 
%%%% S E C T I O N %% III %%%%%%%%%%%%%%%%%%%%%%%%%%%%%%%%%%%%%%%%%%%%%%%%%%%%%%%%%%%%%%%%%%%%%%%%%%%%%%%%%%%%%%%%%%%%%%%%%%%%%%%%%%%%%%%%%%%%%%%%%%%%%%%%%%%%
\section{Visual comparison}\label{comparison}
% Figure 1: fraction of the mass of the most massive object normalized by the stellar mass v/s the star formation efficiency --------------------------------%
\begin{figure*}
    \resizebox{\hsize}{!}
            {
            \centering
            \includegraphics{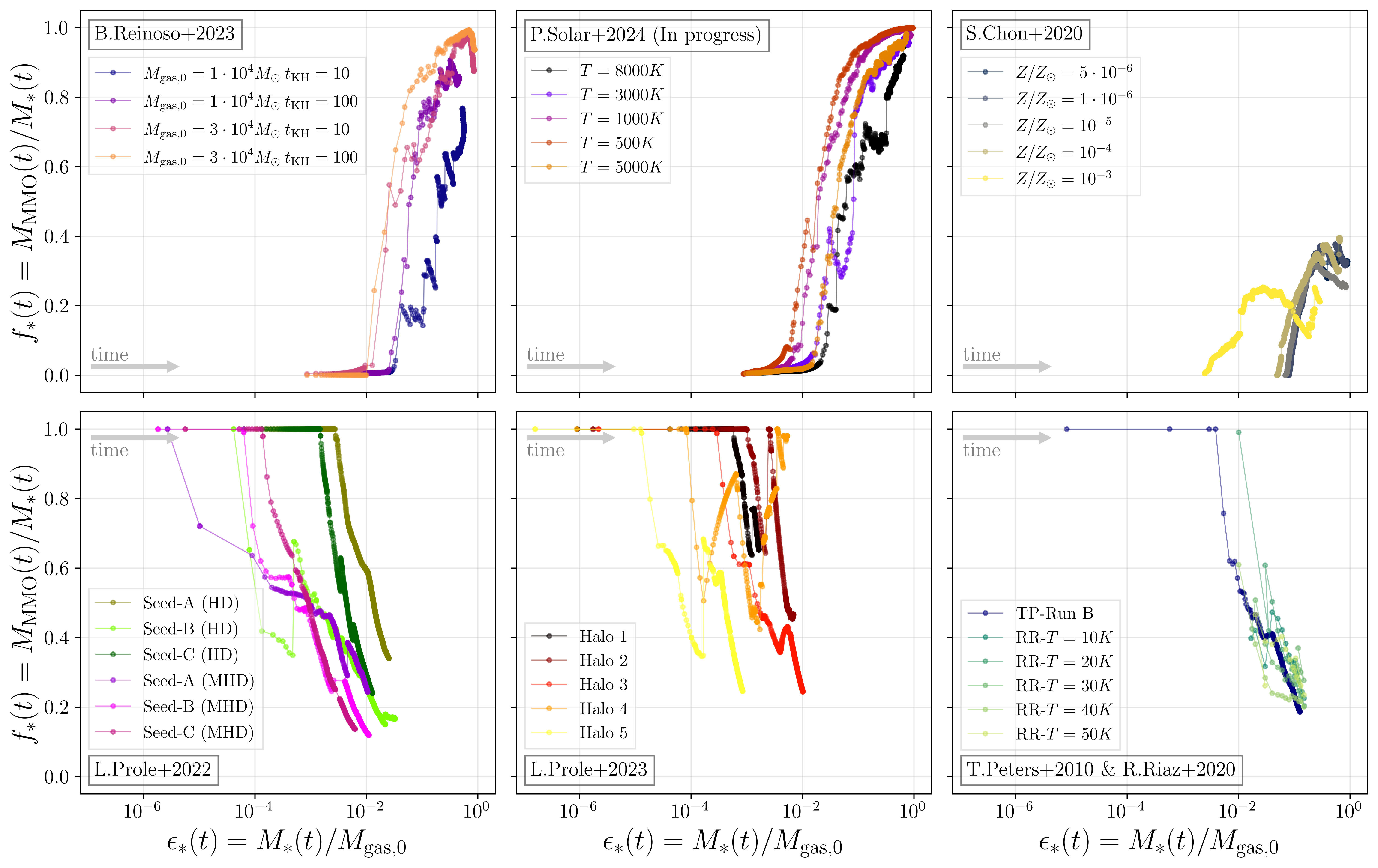}}
    \caption{Fraction $f_*$ of the mass of most massive object $M_{\rm MMO}$ normalized by the total stellar mass $M_*$ shown as a function of the star formation efficiency for the simulations summarized in Table~\ref{tabsims}. The gray arrow indicates the direction of the time evolution. For those authors with more than 8 simulations, we display some representative examples.}
    \label{stellar_vs_sfe}
\end{figure*}
%------------------------------------------------------------------------------------------------------------------------------------------------------------%
% Figure 2: fraction of the mass of the most massive object normalized by the total mass v/s the star formation efficiency ----------------------------------%      
\begin{figure*}
    \resizebox{\hsize}{!}
            {
            \centering
            \includegraphics{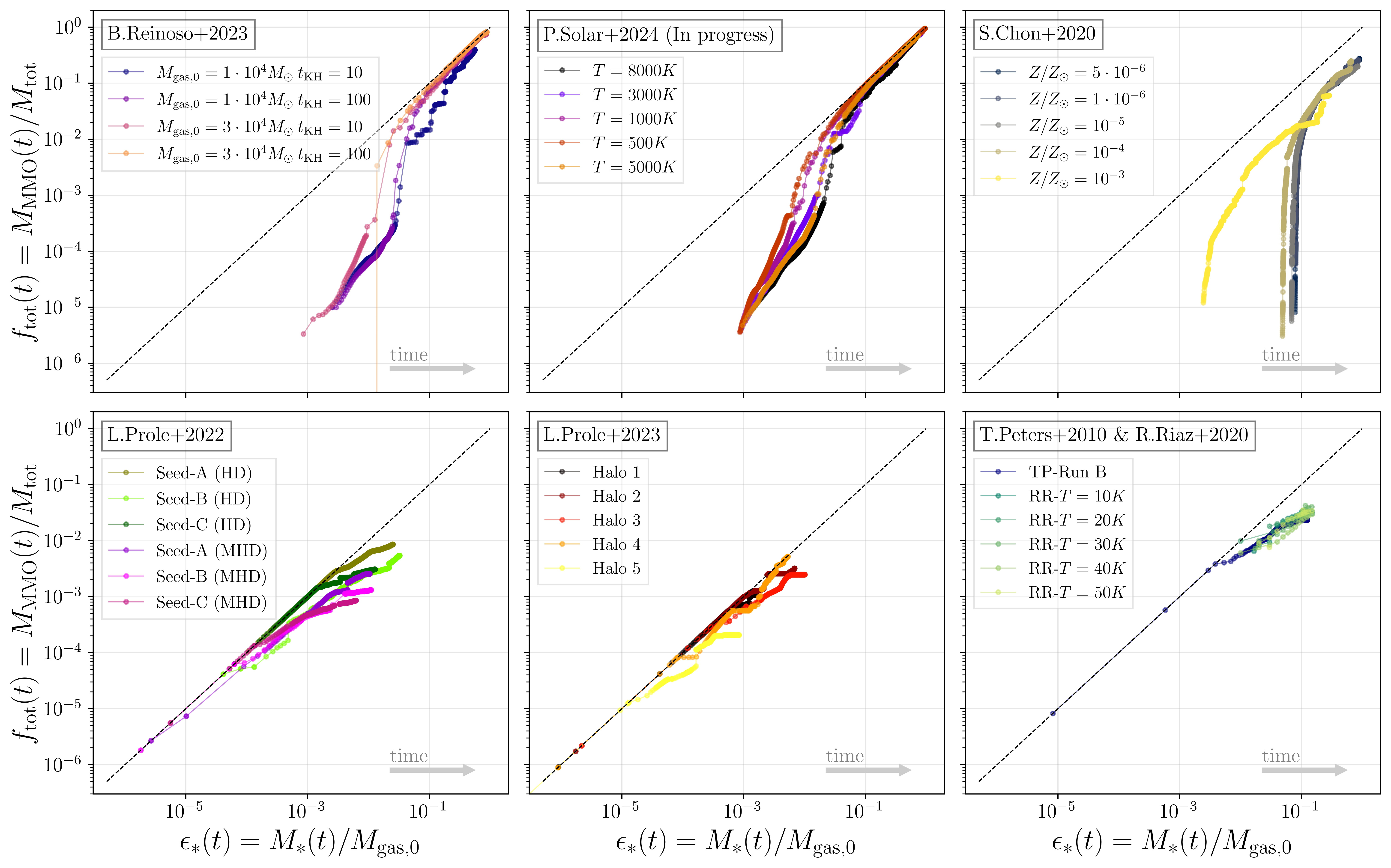}}
    \caption{Fraction $f_{\rm tot}$ of the mass of the most massive object $M_{\rm MMO}$ normalized by the total mass $M_{\rm tot}$ as a function of the star formation efficiency $\epsilon_*$ for the simulations summarized in Table~\ref{tabsims}. The black dotted line corresponds to the identity function, indicating the scenario where all the stellar mass is concentrated within the central object, representing the maximum possible value for $f_{\rm tot}$. The gray arrow indicates the direction of the time evolution. For those authors with more than 8 simulations, we display some representative examples.}
    \label{total_vs_sfe}
\end{figure*}
%------------------------------------------------------------------------------------------------------------------------------------------------------------%
% Figure 3: fraction of the mass of the most massive object normalized by the total mass v/s the collision parameter ----------------------------------------%      
\begin{figure*}
    \resizebox{\hsize}{!}
            {
            \centering
            \includegraphics{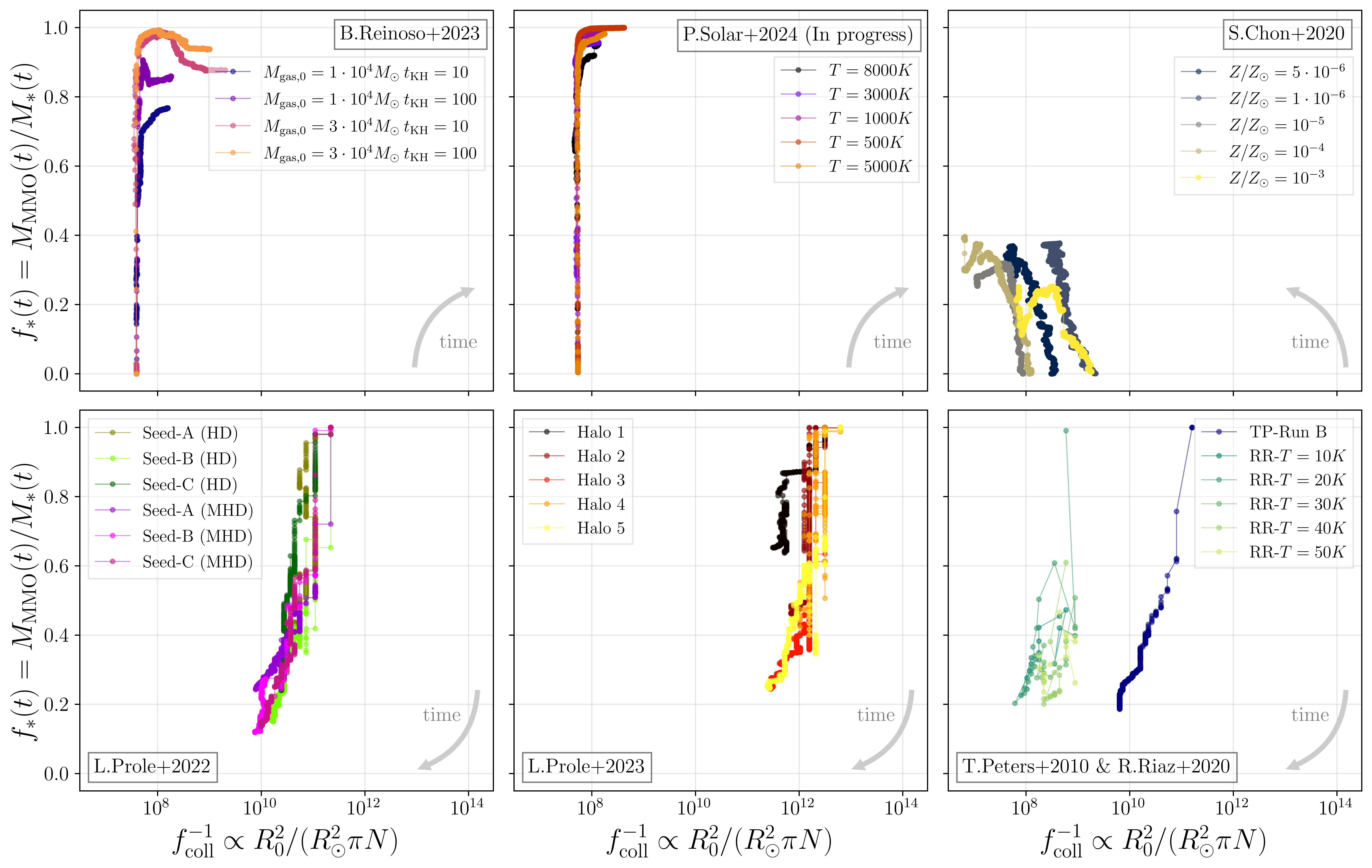}}
    \caption{Fraction $f_*$ of the mass of the most massive object $M_{\rm MMO}$ normalized by the total stellar mass $M_*$ as a function of the collision parameter $f_{\rm coll}^{-1}$ for the simulations summarized in Table~\ref{tabsims}. The gray arrow indicates the direction of the time evolution. For those authors with more than 8 simulations, we display some representative examples.}
    \label{collparam}
\end{figure*}
% -----------------------------------------------------------------------------------------------------------------------------------------------------------%
% 1-About the dimensionless parameters ----------------------------------------------------------------------------------------------------------------------%
As it is clear from the previous section and the summary in Table~\ref{tabsims}, the data are obtained from a range of different situations, with varying cloud masses and metallicities, differences in the physics and the numerical treatment such as the employed integration time after sink particle formation, among others. As we aim to identify similarities between the rather different simulations, a comparison of dimensional numbers is then clearly not meaningful, but it will be important to work with normalized dimensionless quantities to determine how far they govern the behavior that was found in the simulations. While the simulations provide a time evolution, a simple comparison in terms of the time coordinate is also not meaningful, as the characteristic timescales may be different in the different simulations. To characterize the evolutionary stage, we employ the star formation efficiency
% Equation 1: definition for the star formation efficiency --------------------------------------------------------------------------------------------------%
\begin{equation}\label{eps-equation}
    \epsilon_*(t)=\frac{M_*(t)}{M_{\rm gas,0}},
\end{equation}
%------------------------------------------------------------------------------------------------------------------------------------------------------------%
where $M_*(t)$ is the total stellar mass at time $t$ and $M_{\rm gas,0}$ is the initial gas mass available to form stars. We further consider the fraction of the total stellar mass $f_*(t)$ that corresponds to the mass of the most massive object $M_{\rm MMO}(t)$ at time $t$, defined via
% Equation 2: definition for the fraction of the mass for the most massive object normalized by the stellar mass --------------------------------------------% 
\begin{equation}\label{fstar-equation}
    f_*(t)=\frac{M_{\rm MMO}(t)}{M_*(t)},
\end{equation}
%------------------------------------------------------------------------------------------------------------------------------------------------------------% 
as well as the fraction of the total mass $f_{\rm tot}$ that corresponds to the most massive object, 
% Equation 3: definition for the fraction of the mass for the most massive object normalized by the total mass ----------------------------------------------% 
\begin{equation}\label{ftot-equation}
    f_{\rm tot}(t)=\frac{M_{\rm MMO}(t)}{M_{\rm tot}},
\end{equation}
%------------------------------------------------------------------------------------------------------------------------------------------------------------% 
where $M_{\rm tot}$ is the sum of the initial mass of the gas and the stellar mass available in the cloud, i.e.
% Equation 4: definition for the total mass -----------------------------------------------------------------------------------------------------------------% 
\begin{equation}
    M_{\rm tot}=M_{\rm gas,0}+M_*(t=0).
\end{equation}
%------------------------------------------------------------------------------------------------------------------------------------------------------------% 
% 2-Explanation about Fig.1 ---------------------------------------------------------------------------------------------------------------------------------%      
In Fig.~\ref{stellar_vs_sfe}, we plot $f_*$ as a function of $\epsilon_*$ for the set of different simulations in Table~\ref{tabsims}. From this plot, we note that the simulations by \citet{Reinoso2023}, Solar et al. (2024-in progress) and \citet{Chon2020} reach star formation efficiencies of $40-90\%$. The simulations by \citet{Peters2010} and \citet{Riaz2020} reach star formation efficiencies of the order $10\%$, while \citet{Prole2022, Prole2023}  reach efficiencies of the order of a few percent. We note here that these differences are often determined by the available runtime of the simulation, the numerical methodology, and the general objective of the simulation. However, also physical processes can play a relevant role; for example, towards the end of the simulations by \citet{Peters2010}, the HII region starts to expand, shutting off accretion and thus limiting the further growth of the star formation efficiency. Other simulations have not yet reached the stage, where feedback is relevant and their efficiencies are rather limited by the runtime of the simulations. For the functional dependence of $f_*$ on $\epsilon_*$, we can identify two main types of behaviors in Fig.~\ref{stellar_vs_sfe}: In the simulations by \citet{Reinoso2023}, Solar et al. (2024-in progress) and \citet{Chon2020}, $f_*$ is initially low and subsequently increases, implying that the time evolution corresponds to a shift towards higher star formation efficiencies and higher fractions $f_*$. In the simulations by \citet{Reinoso2023} and Solar et al. (2024-in progress), this is a direct result of the initial conditions which include $256$ protostars of equal mass, implying a low initial value of $f_*$ that subsequently increases as a result of accretion and mergers. Similarly, in the simulations of \citet{Chon2020}, a significant number of sink particles forms very early, leading to a low initial value of $f_*$.

On the other hand, in the simulations by \citet{Prole2022b, Prole2023}, \citet{Peters2010} and \citet{Riaz2020}, sink particle formation occurs more gradually, starting initially with one sink particle and thus implying an initial value $f_*=1$, which subsequently decreases as additional sink particles form and acquire mass. We further see that the simulations terminate in rather different regimes; while in the simulations of \citet{Reinoso2023}, Solar et al. (2024-in progress) and \citet{Chon2020}, a very large fraction of the stellar mass is in the most massive particle ($35-99.7\%$), the latter is more moderate in the simulations by \citet{Prole2022b, Prole2023}, \citet{Peters2010} and \citet{Riaz2020} with more typical values in the range of $20-30\%$. 

The difference in this behavior in principle can be explained as a result of the initial conditions as well as the time evolution covered in the simulations. For example, as the simulations of \citet{Reinoso2023} and Solar et al. (2024-in progress) already start with initial protostars, the initial star formation efficiency in these simulations is already $\sim0.25\%$, while most of the evolution in the models by \citet{Prole2022b, Prole2023} covers lower efficiencies and different phases of the star formation process. Indeed, the simulation by \citet{Chon2020} suggests that the situation may be more complex, including phases where $f_*$ increases with $\epsilon_*$ and others where $f_*$ decreases with increasing $\epsilon_*$. Nonetheless, some of the differences observed here could be a result of the rather different environments, an aspect we will consider in more detail below.

% 3-Explanation about Fig.2 ---------------------------------------------------------------------------------------------------------------------------------%      
Now, we consider the evolution of $f_{\rm tot}$ as a function of $\epsilon_*$ in Fig.~\ref{total_vs_sfe}. While there are evident differences in the range of the star formation efficiencies covered by the different simulations as discussed in the context of Fig.~\ref{stellar_vs_sfe}, the evolution overall appears here more similar, as $f_{\rm tot}$ is generally found to increase as a function of $\epsilon_*$ with at most minor and temporary deviations. Differences are however visible in the slope of the increase, with the simulations by \citet{Reinoso2023}, Solar et al. (2024-in progress) and \citet{Chon2020} increasing more steeply, and a more gradual increase in the simulations by \citet{Prole2022b}, \citet{Prole2023}, \citet{Peters2010} and \citet{Riaz2020}. In the simulations by \citet{Reinoso2023}, Solar et al. (2024-in progress) and \citet{Chon2020}, fractions $f_{\rm tot}$ of $30-100\%$ are being reached, while this value remains in the percent or sub-percent range in the simulation of \citet{Prole2022b, Prole2023}, \citet{Peters2010} and \citet{Riaz2020}, also reflecting the difference in the evolutionary stage as defined by the star formation efficiency $\epsilon_*$. 

% 4-About the collisions in the simulations -----------------------------------------------------------------------------------------------------------------%      
While most of the simulations we compare in this study (except \citet{Peters2010})  allow for mergers to happen, we note that these are particularly frequent in the simulations of \citet{Reinoso2023}, Solar et al. (2024-in progress) and \citet{Chon2020}. It is thus conceivable that some of the major differences we find in the simulations, particularly concerning the obtainable values of $f_*$, are driven by the possibility to have collisions within the systems. In the following, we derive a simple criterion to estimate whether collisions can be expected to be relevant in a given simulation. Given an initial cloud mass $M_{\rm gas,0}$ and an average mass per protostar $m_*$, the number of protostars can be estimated to be
% Equation 5: Estimation for the number of protostars -------------------------------------------------------------------------------------------------------%      
\begin{equation}\label{nfrac-equation}
    N=\frac{\epsilon_*M_{\rm gas,0}}{m_*}.
\end{equation}
% -----------------------------------------------------------------------------------------------------------------------------------------------------------%
The number density of the protostars follows as $n=N/V$, where the volume $V$ is calculated from the radius $R_0$ assuming spherical symmetry, i.e. $V=\frac{4\pi}{3}R_{0}^3$. Estimating the collision length $\lambda$ via the mean free path, we have
% Equation 6: Definition of collision length ----------------------------------------------------------------------------------------------------------------%      
\begin{equation}\label{lambda-equation}
    \lambda=\frac{1}{n\sigma},
\end{equation}
% -----------------------------------------------------------------------------------------------------------------------------------------------------------%
where $\sigma$ is the effective cross section for collisions. Comparing the collision length to the radius of the cloud and inserting the above expressions, we find
% Equation 7: Definition of collision parameter -------------------------------------------------------------------------------------------------------------%      
\begin{equation}\label{fcol-equation}
    f_{\rm coll}^{-1}=\frac{\lambda}{R_0}\sim \frac{R_0^2}{\epsilon_* \sigma (M_{\rm gas,0}/m_*)}\sim \frac{R_0^2}{\sigma N}.
\end{equation}
% -----------------------------------------------------------------------------------------------------------------------------------------------------------%
In our estimate, we aim not to anticipate any particular properties of the protostars that form but rather to assess very conservatively if collisions might be occurring within the system. Therefore, we conservatively estimate $\sigma\sim R_\odot^2\pi$, with $R_\odot$ the solar radius, noting that of course this likely represents an underestimate for simulations in which very massive objects are forming. While the information incorporated in this parameter is not complete, a decreasing value of $f_{\rm coll}^{-1}$ may then suggest more frequent collisions.
 
% 5-Explanation about Fig.3 ---------------------------------------------------------------------------------------------------------------------------------%      
We now present the fraction $f_*$ of the mass of the most massive object normalized by the total stellar mass $M_*$ as a function of the collision parameter $f_{\rm coll}^{-1}$ in Fig.~\ref{collparam}. We consider both parameters as a function of time, particularly accounting for the time evolution of the number of protostars $N$ that have formed at time $t$. We find that the simulations of \citet{Reinoso2023}, Solar et al. (2024-in progress) and \citet{Chon2020} have systematically lower values of $f_{\rm coll}^{-1}$ compared to the other runs. Particularly, \citet{Reinoso2023} and Solar et al. (2024-in progress) report initial values $f_{\rm coll}^{-1}\sim10^{8}$, shifting towards $f_{\rm coll}^{-1}\sim10^{9}$ when a significant fraction of the mass concentrates in the most massive object, indicating efficient collisions to occur. For the simulations of \citet{Chon2020}, we have a range of $f_{\rm coll}^{-1}\sim10^{7}-10^{9.5}$, where the collision parameter increases over time as new protostars continue to form very efficiently. In the simulations by \citet{Prole2022b}, the collision parameter starts out at $f_{\rm coll}^{-1}\sim10^{10}$ and shifts towards higher values of $f_{\rm coll}^{-1}\sim3\times10^{11}$ as a result of some collisions occurring, but more gradually over time. The simulations by \citet{Prole2023} represent even less evolution in this parameter and for most of it remain close to $f_{\rm coll}^{-1}\sim10^{12}$. The \citet{Peters2010} simulations start out similarly as \citet{Prole2022b} from $f_{\rm coll}^{-1}\sim10^{10}$ subsequently and gradually evolving towards higher values. The simulations by \citet{Riaz2020} start out at $f_{\rm coll}^{-1}\sim10^{8}$ but move towards higher values already when $f_*$ is still low, indicating some mergers rather in the earlier phase of the evolution, and more distributed in time compared to the situation modeled by \citet{Reinoso2023} or Solar et al. (2024-in progress). It is conceivable that the impact after the first mergers in the simulations of \citet{Riaz2020} on the collision parameter is more significant, considering the overall lower number of protostars in the respective runs.

%%%% S E C T I O N %% IV %%%%%%%%%%%%%%%%%%%%%%%%%%%%%%%%%%%%%%%%%%%%%%%%%%%%%%%%%%%%%%%%%%%%%%%%%%%%%%%%%%%%%%%%%%%%%%%%%%%%%%%%%%%%%%%%%%%%%%%%%%%%%%%%%%%%%      
\section{Machine learning analysis}\label{ML}
% Table 2: Summary of the metrics for different models ------------------------------------------------------------------------------------------------------%  
\begin{table*}[!t]
    \centering
    \begin{tabular}{llccc}\hline
         {\bf F.Extraction}         & {\bf ML-Model} & {\bf $R^2$}      & {\bf MAE}        & {\bf RMSE}         \\ \hline
         -                          & SVM            &  $0.479\pm0.007$ &  $0.182\pm0.002$ & $0.214\pm0.002$  \\
         Standard scaling           & Random Forest  &  \textbf{0.942} $\pm$ \textbf{0.002} &  \textbf{0.029} $\pm$ \textbf{0.001} & \textbf{0.070} $\pm$ \textbf{0.001} \\
         $t$-test feature selection & XGBoost        &  $0.930\pm0.001$ &  $0.047\pm0.007$ & $0.078\pm0.001$  \\
         $t$-test feature selection & LightGBM       &  $0.928\pm0.002$ &  $0.039\pm0.001$ & $0.079\pm0.002$  \\
         -                          & k-NN           &  $0.714\pm0.004$ &  $0.102\pm0.001$ & $0.158\pm0.001$  \\ \hline
    \end{tabular}
    \caption{Summary of the test stage for the different models obtained in the workflow. In addition to RMSE we include additional metrics to evaluate the results, the coefficient of determination ($R^2$-score) and Mean Absolute Error (MAE). Both RMSE and MAE are minimized to 0, while the $R^2$-score is maximized to 1 (i.e. the predictions are perfect). Note that the best ML method is bold faced.}\label{tabmetrics}
\end{table*}
%------------------------------------------------------------------------------------------------------------------------------------------------------------%
% 1-About the importance and help than ML can provide in these study ----------------------------------------------------------------------------------------%  
After visually comparing the simulations employed in this study, we can conclude that different environments in the simulations lead to different relationships between $\epsilon_{*}$ and $f_{*}(t)$. In this section, we aim to explore whether we can extract relevant information from these simulations using machine learning (ML) models to arrive at accurate parameter estimations based on the input from these simulations. In principle, these models may account for highly nonlinear behaviors \citep[e.g.,][]{Bishop2006}.
% 2-Formalization of the problem ----------------------------------------------------------------------------------------------------------------------------%
One of the objectives of these simulations is to study the formation and evolution of the most massive object. It is our interest to investigate whether an ML model ($f_{\vec{\theta}}$) of parameters $\vec{\theta}$ is able to estimate $f_{*}(t_i)$ given $n$ attributes or features $\vec{x}_i = (x_{i,1}, x_{i,2}, ..., x_{i,n})$ related to the specific time $t_i$ of snapshot $i$ since the formation of the first sink particle in the simulation. Considering our dataset $\mathcal{D}=\{(\vec{x}_1, y_1), (\vec{x}_2, y_2),...,(\vec{x}_q, y_q)\}$, comprised of $q$ pairs of attributes $\vec{x}_i\in\mathds{R}^n$ and labels $y_i\in\mathds{R}$ obtained from all the simulations, the regression problem can be defined as:
% Equation 8: The regression problem-------------------------------------------------------------------------------------------------------------------------%
\begin{equation}\label{reg-equation}
    f_{\vec{\theta
    }}(\vec{x}_i)=\hat{y}_{i}\rightarrow \hat{y}_{i}\approx y_{i},
\end{equation}
% -----------------------------------------------------------------------------------------------------------------------------------------------------------%
where $\hat{y}_i$ corresponds to the prediction of the ML model, and $y_{i}=f_{*}(t_i)$. We seek to adjust $\vec{\theta}$ so that $f_{\vec{\theta}}$, given a cost function $\mathcal{L}(\hat{y}_{i},y_{i})$, can make accurate predictions about the mass of the most massive object, independent of the type of regime that dominates the gravitational collapse of the cloud.

We will proceed to explain the preprocessing performed on our dataset, define the attributes that will be used as input for the model, discuss the model selection process, and finally, we will present the results and predictions obtained.
% 3-Data preprocessing --------------------------------------------------------------------------------------------------------------------------------------%  
\subsection{Data preprocessing}
% Figure 4: Spearman correlogram plot -----------------------------------------------------------------------------------------------------------------------%
\begin{figure}
    \resizebox{\hsize}{!}
            {
            \centering
            \includegraphics{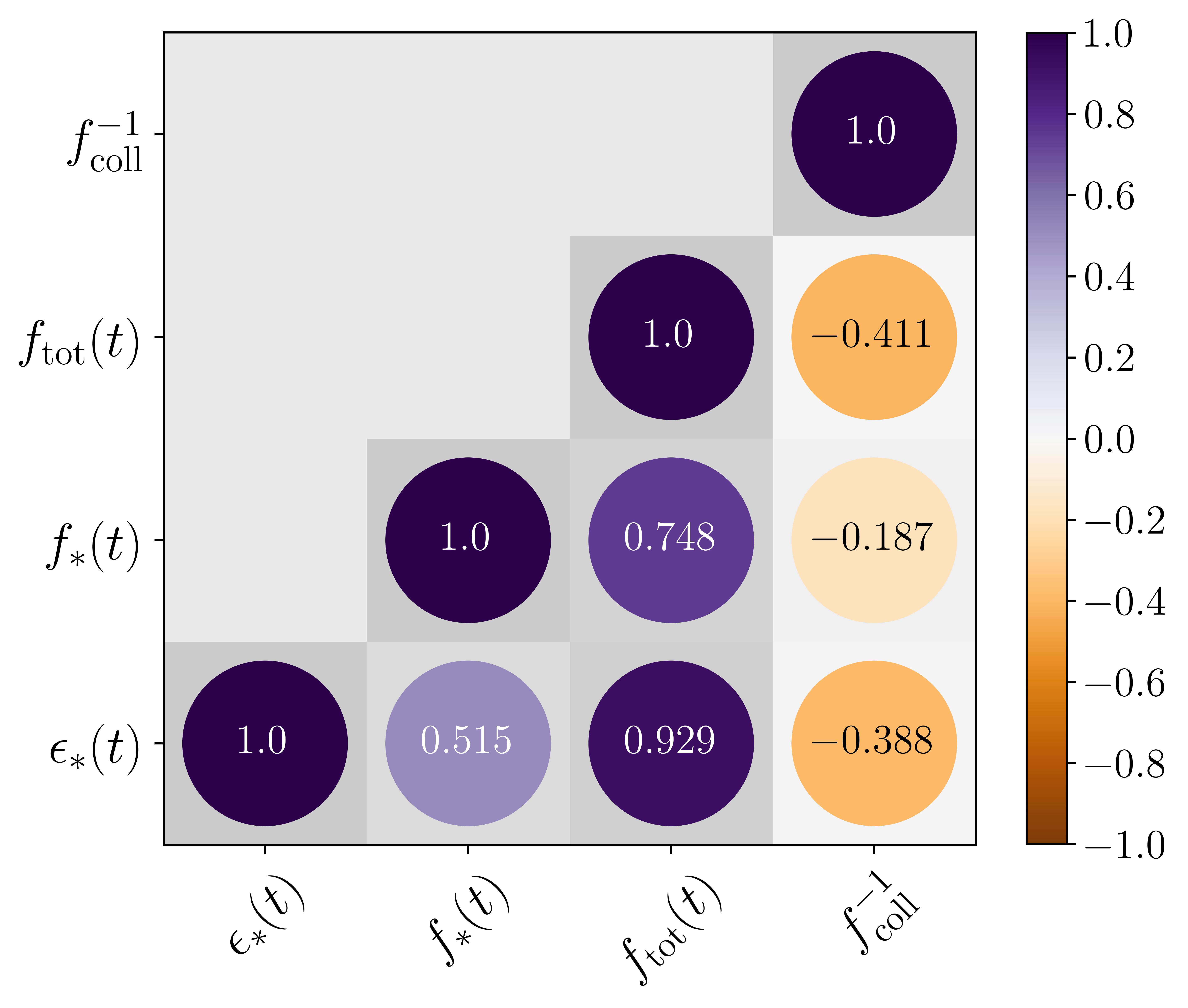}}
    \caption{Correlogram for the dimensionless parameters calculated in this work. The values shown correspond to Spearman's correlation coefficient $\rho_{s}$, which assesses the monotonic relationship between the ranks of the paired data points. }
    \label{spearmancorr}
\end{figure}
%------------------------------------------------------------------------------------------------------------------------------------------------------------%
Although we successfully assembled a sufficiently large dataset covering various environments, the simulations span different ranges of $\epsilon_{*}(t)$, and the time step for calculating each snapshot varies. These characteristics of the data may lead to overfitting during training or introduce bias towards runs with a higher number of points. To mitigate these potential issues, we initially limit the range of $\epsilon_{*}(t)$ to values between 0 and 0.4. Additionally, we employ linear interpolation for each run, ranging from its minimum to its maximum value, resulting in new curves that are 500 points in length. 

Now, concerning the variables that will be supplied to the model for estimating $f_{*}(t_i)$, we regard $\epsilon_{*}(t_i)$ as a key parameter, offering insights into the amount of gas mass utilized in protostar formation. Similarly, as previously discussed, mergers play a crucial role in comprehensively characterizing the simulation environment. Therefore, we will incorporate $f_{\rm coll}^{-1}(t_i)$ as another significant factor. 

To analyze if the dimensionless parameters we have chosen can effectively describe $f_{*}(t)$, we calculated their correlation coefficients. In Fig.~\ref{spearmancorr}, we use the Spearman correlation coefficient to analyze the relationship between the dimensionless parameters $\epsilon_{*}$, $f_{*}$, $f_{\rm tot}$, and $f_{\rm coll}^{-1}$ for the data in our sample. We find particularly strong correlations between $f_{\rm tot}$ and $\epsilon_{*}$ ($0.929$), $f_{\rm tot}$ and $f_{*}$ ($0.748$), and a moderate correlation between $\epsilon_{*}$ and $f_{*}$ ($0.515$). Of course, this is not fully unexpected, as we have $f_{\rm tot}\approx\epsilon_*f_*$; however, the analysis interestingly shows that the correlation is particularly strong between $f_{\rm tot}$ and $\epsilon_*$, while still strong but to a lesser degree between $f_{*}$ and $\epsilon_*$. Additionally, a moderate anti-correlation is present between $f_{\rm tot}$ and $f_{\rm coll}^{-1}$ ($-0.411$) and between $\epsilon_{*}$ and $f_{\rm coll}^{-1}$ ($-0.388$). From this calculation, we can conclude that the combination of $f_{\rm coll}^{-1}$ and $\epsilon_{*}$ forms pairs with different types of correlations with $f_{*}$ and with each other, making them suitable candidates as input values for the ML model. We will not take into consideration the $f_{\rm tot}$ parameter as an attribute for the training of the models since, as can be seen in Eq.~\ref{ftot-equation}, it is calculated directly with the value of the mass of the most massive object.

Finally, to mitigate overfitting and enhance confidence in predictions, we randomly divided our dataset into three subsets: training (16,640 values), validation (4,160 values), and testing (5,200 values). The training of the best candidates obtained were repeated in a 3-fold manner, with each iteration employing a distinct random data partitioning, ensuring the robustness of the predictions.

% 4-Model Selection -----------------------------------------------------------------------------------------------------------------------------------------%
\subsection{Model selection}
The selection of the most suitable ML model is a computationally expensive task, often influenced by the nature and dimensionality of the data. To ensure that we identify the best model within a range of possibilities, we execute an \texttt{AutoML} workflow from the \textit{Modulos AI}\footnote{AutoML: https://www.modulos.ai/} platform, on the dataset for training and validation. The \texttt{AutoML} platform (version: 1.1.2) is designed for automated model selection and supervised training in a given task. This process involves detecting the uploaded data type, defining the task's objective, and optimizing the combination of feature extractors, machine learning algorithms, and hyperparameters using a Bayesian optimizer \citep{Srinivas2009}. The platform identifies a candidate, trains it, validates it, and delivers the solution to the user. This cycle repeats until reaching a user-defined endpoint..

\texttt{AutoML} also can evaluate the efficiency of different preprocessing techniques or feature extractors before training the models, or it can leave the data as it is (referred to as the identity transformation on the platform). The preprocessing options that we will study include standard scaling, which transforms the features of a dataset to have a mean of 0 and a standard deviation of 1, and $t$-test feature selection. The latter involves dividing the data into groups, applying the $t$-test to each feature, and selecting features with statistically significant differences in their means \citep{Zhou2007}.

Regarding the possible learning algorithms in the platform, we select:  
\begin{itemize}
    \item \textit{k-Nearest Neighbors} (kNN): Operating on the principle of proximity in feature space, kNN identifies the $k$-nearest data points to a given instance based on a predefined metric. In regression, the algorithm predicts the target value by averaging or aggregating the values of these nearest neighbors. The parameter $k$ delineates the number of neighbors considered, acting as a decision boundary for data distribution \citep{Fix1989, Altman1992}.
    
    \item \textit{Support Vector Machine} (SVM): The main idea consists of finding a hyperplane that maximizes the margin between data points and a regression line. For the non-linear case, a kernel function is used to map the data. SVMs minimize prediction error within a loss function, allowing for a tolerance region around the regression line \citep{Cortes1995}. 
    
    \item \textit{Random Forest} (RF): A robust algorithm that combines multiple decision trees for enhanced predictive power. Utilizing bootstrap aggregating, random subspace, and decision trees \citep{Ho1998}, it draws bootstrap samples, constructs decision trees, and employs majority voting for final predictions \citep{Breiman2001}. 
    
    \item \textit{Extreme Gradient Boosting} (XGBoost): An efficient and high-performance supervised learning method, rooted in the gradient boosting algorithm \citep{Friedman2001GBoost}. It operates by sequentially incorporating decision trees into its model, selecting trees that minimize the gradients between the target variable and the model's predictions at each step. This process continues until a predetermined number of iterations or a specified error threshold is reached \citep{Chen2016}.
    
    \item \textit{LightGBM}: Also with a gradient boosting framework, LightGBM employs a tree-based learning strategy, similar to XGBoost, and distinguishes itself through its focus on leaf-wise tree growth, where each added node corresponds to the leaf with the maximum delta loss, resulting in a more complex but narrower tree structure \citep{Ke2017}. 
\end{itemize}

To evaluate the confidence of the models, we take the root mean square error (RMSE) as the cost function:
% Equation 8: RMSE -----------------------------------------------------------------------------------------------------------------------------------------------%    
\begin{equation}
    \text{RMSE} = \sqrt{\frac{1}{n}\sum_{i=1}^{n}(y_{true,i}-y_{pred,i})^2},
\end{equation}
%-----------------------------------------------------------------------------------------------------------------------------------------------------------------%
which is widely utilized in machine learning algorithms, particularly regression models, to quantify the accuracy of predictions. It gauges the square root of the average squared differences between predicted and actual values, providing a comprehensive measure of the model's predictive performance. 

% 5-Results -------------------------------------------------------------------------------------------------------------------------------------------------%
\subsection{Results}
% Figure 5: Author palete correlation plot for the best model -----------------------------------------------------------------------------------------------%
\begin{figure}
    \resizebox{\hsize}{!}
            {
            \centering
            \includegraphics{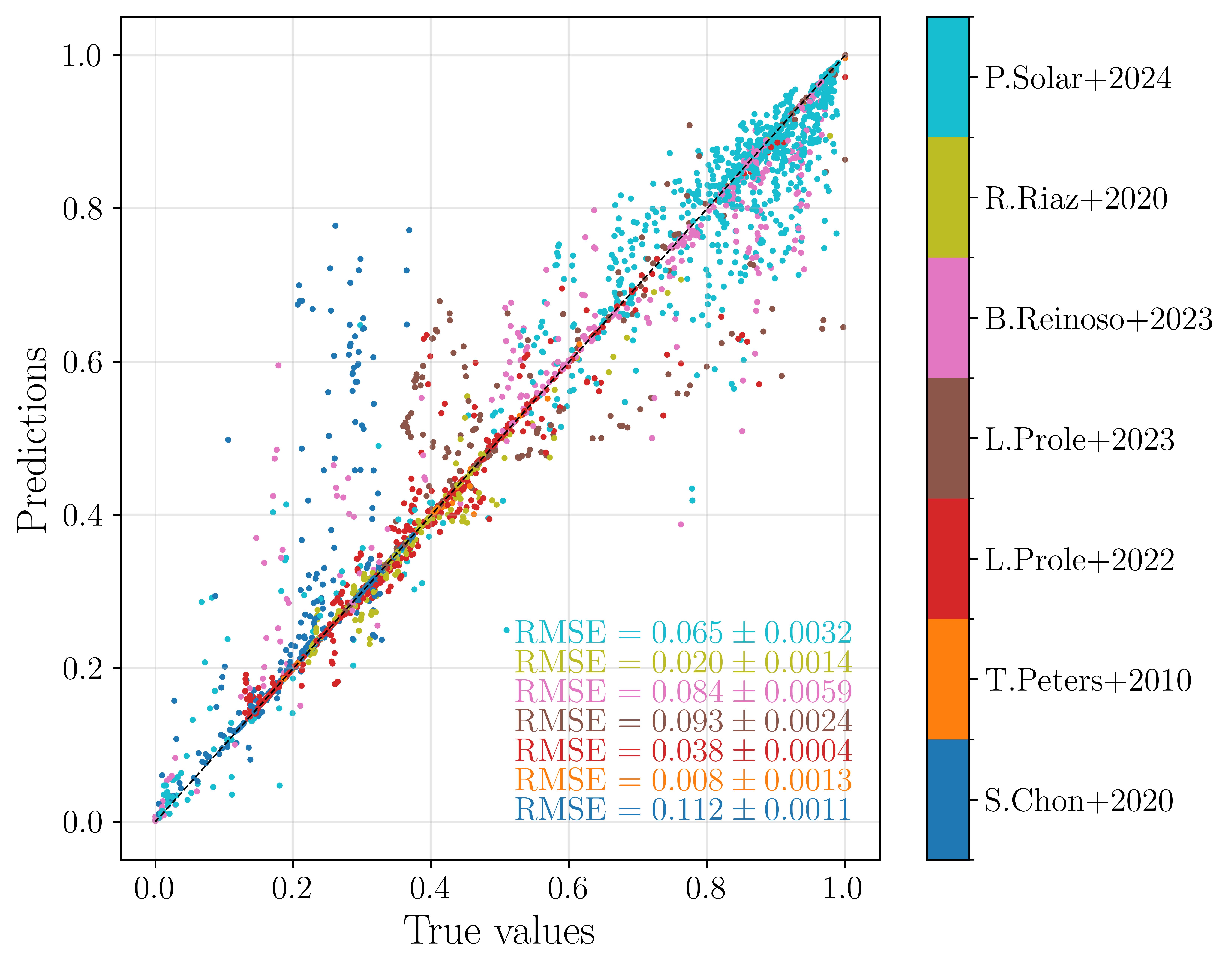}}
    \caption{Correlation plot for the average of the $f_{*}(t)$ predictions of the test set by RF. The color palette represents the different authors of the simulations, and we include the RMSE for each case.}\label{corrplot}
\end{figure}
%------------------------------------------------------------------------------------------------------------------------------------------------------------%
We execute the workflow described above with the intention that if there are no improvements in the RMSE score using the validation set after 350 solution candidates, it is paused. We select the best candidate for each algorithm, retrain it in a 3-fold manner from scratch, and utilize the test set to evaluate its performance (for more information, see Appendix~\ref{AppendixA}). These results can be seen in Table~\ref{tabmetrics}, where we summarize different metrics besides RMSE for completeness. Simple algorithms such as kNN and SVM fail to predict with confidence, yielding lower results on each different metric. This can be attributed to a variety of reasons, but looking at the interactions between the dimensionless parameters in Fig.'s~1-3, they show a high scatter for different authors, even after preprocessing, which strongly affects kNN and SVM, as they are sensitive to high variances in the data.

In contrast, ensemble models, such as RF, LightGBM, and XGBoost, exhibit robustness against outliers, thereby enabling them to generate predictions with heightened reliability. Notably, RF distinguishes itself by demonstrating superior performance across all evaluation metrics. This efficacy is principally attributable to the robustness that RF has against noisy data, which lets it handle irrelevant features without significantly impacting its performance. Consequently, RF emerges as the preferred  model for our subsequent analysis.

\begin{figure}
    \resizebox{\hsize}{!}
            {
            \centering
            \includegraphics[width=0.85\columnwidth]{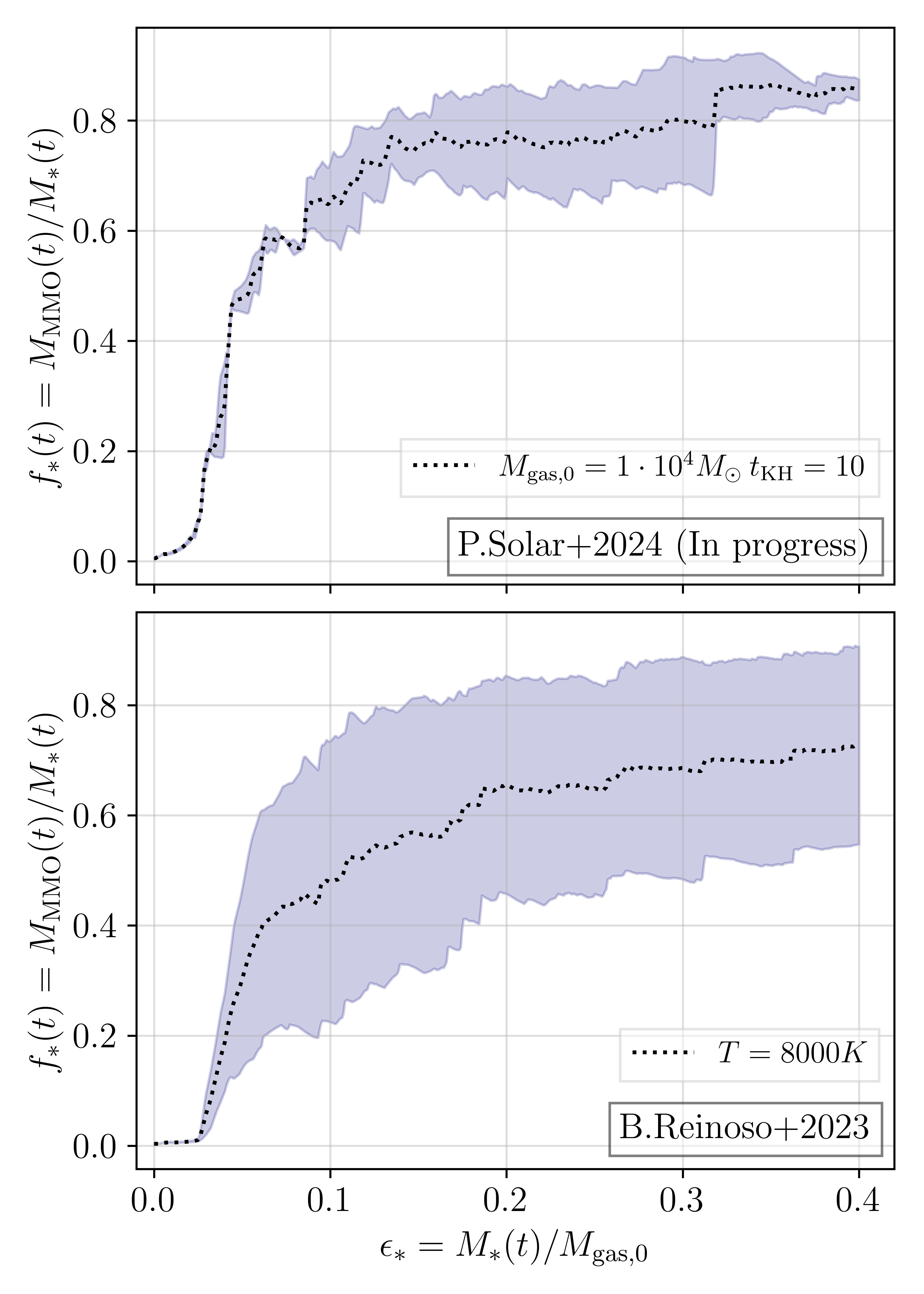}}
    \caption{Examples of the dispersion of $f_{*}$ in simulations of Solar et al. (2024-in progress) (2 runs, top panel) and \citet{Reinoso2023} (6 runs, bottom panel) with the same initial conditions but different seeds. The dashed line corresponds to the average between the simulations and the shading to the standard deviation at each point of $\epsilon_{*}$.}\label{simsstd}
\end{figure}

\begin{figure*}
    \resizebox{\hsize}{!}
            {
            \centering
            \includegraphics{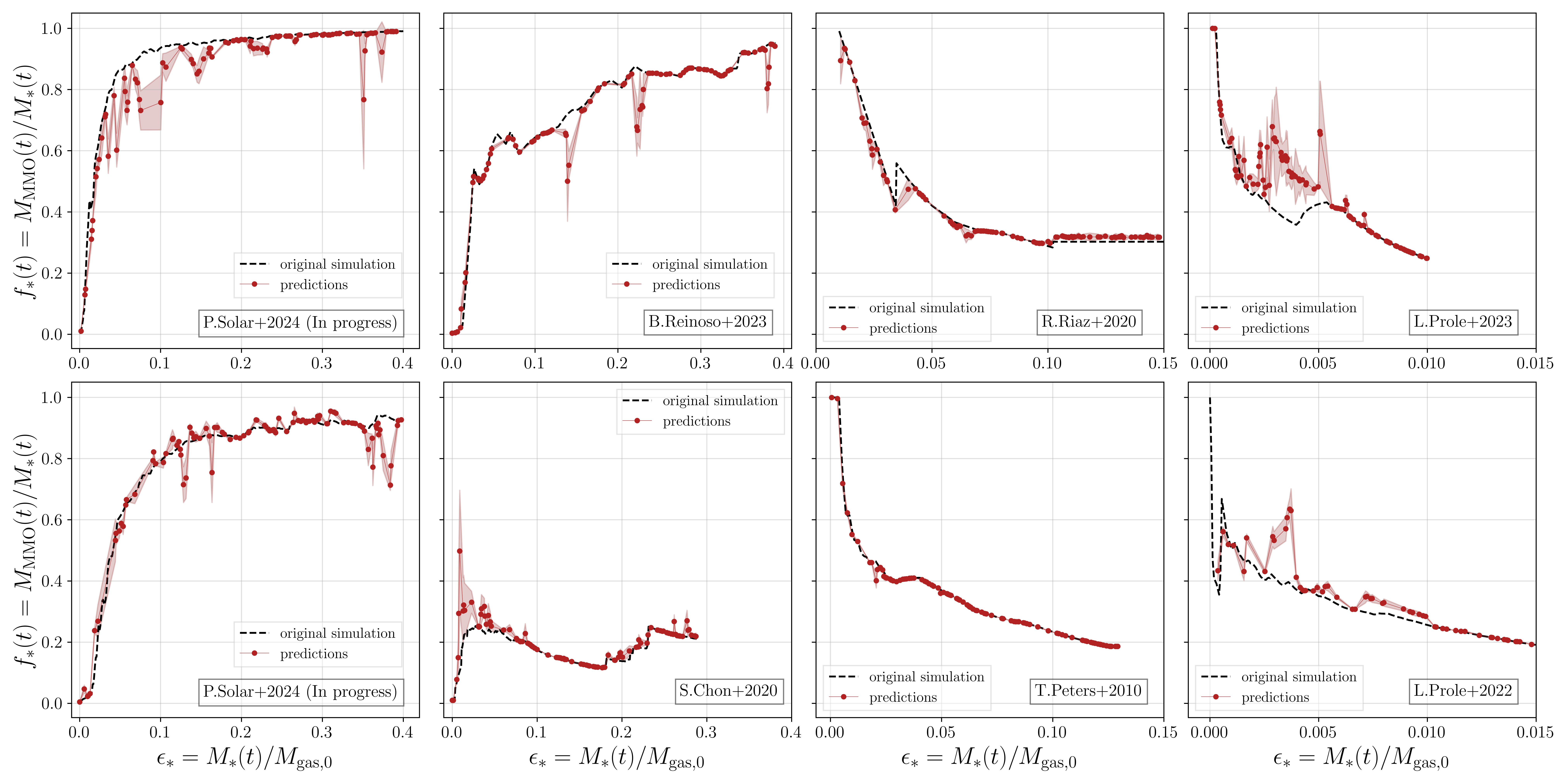}}
    \caption{Predicted values for $f_{*}(t)$ by RF are depicted. The dashed black line represents the values of the simulation, the red dots indicate the average of the predictions, and the shaded area represents the standard deviation of predictions. We present random examples for all authors and for different ranges of $\epsilon_{*}(t)$.}
    \label{ML-sample}
\end{figure*}

A comparison between original and predicted values of $f_{*}(t)$ is given in Fig.~\ref{corrplot} for the different simulations in our test sample. We generally find that most of the predictions are close to the true values so the majority of the data points are close to the diagonal lines of the diagram. Particularly good predictions are obtained for the simulations of \citet{Peters2010}, \citet{Riaz2020}, and \citet{Prole2022b}. Some of the predicted values for the simulations of \citet{Chon2020}, Solar et al. (2024-in progress), and \citet{Reinoso2023} lie further away from the simulated data points. The latter can be attributed to the high dispersion of $f_{*}$ in simulations sharing the same initial conditions, as it can be misinterpreted by the models as noisy data. An example of this can be seen in Fig.~\ref{simsstd}, where we calculate the average and standard deviation of $f_{*}$ for two different authors. For illustration, we provide typical examples of the reconstruction of the relation between $f_{*}$ and $\epsilon_{*}$ in Fig.~\ref{ML-sample}, showing that the predictions of $f_*(t)$ obtained via Random Forest are relatively close to the original simulations. 
%------------------------------------------------------------------------------------------------------------------------------------------------------------%
% Figure 6: Standard deviation of the simulations -----------------------------------------------------------------------------------------------------------%
%------------------------------------------------------------------------------------------------------------------------------------------------------------%
% Figure 7: Sample reconstruction using predicted points by the ML-Model ------------------------------------------------------------------------------------%
%------------------------------------------------------------------------------------------------------------------------------------------------------------%

%%%% S E C T I O N %% V %%%%%%%%%%%%%%%%%%%%%%%%%%%%%%%%%%%%%%%%%%%%%%%%%%%%%%%%%%%%%%%%%%%%%%%%%%%%%%%%%%%%%%%%%%%%%%%%%%%%%%%%%%%%%%%%%%%%%%%%%%%%%%%%%%%%%%
\section{Summary}\label{discussion}
% 1- Main ideas of the study --------------------------------------------------------------------------------------------------------------------------------%
The formation of massive objects via gravitational collapse is important to explain the origin of massive black holes and stars. Here we have analyzed and compared numerical simulations of gravitational collapse in different environments, {focusing on the regime of low initial ratios of turbulent to gravitational energy, low to moderate Mach numbers and the absence of strong feedback. We consider} the formation of massive black holes in the early Universe \citep{Chon2020, Solar2022, Reinoso2023}, the formation of primordial massive stars \citep{Prole2022b, Prole2023}, star formation at low metallicity \citep{Riaz2020} and formation of massive stars in molecular cores \citep{Peters2010}. {As most of these simulations employ a piecewise-polytropic equation of state, our dataset does not allow to analyse the effect of the cooling time on the dynamics or the evolution of the most massive object. }
%------------------------------------------------------------------------------------------------------------------------------------------------------------%
% 2- Methodology and results --------------------------------------------------------------------------------------------------------------------------------%
For a meaningful comparison of the simulations of different environments, we have introduced dimensionless parameters, particularly the star formation efficiency $\epsilon_{*}$ describing the fraction of mass that was turned into stars, the ratio $f_{*}$ corresponding to the mass of the most massive star normalized by the total stellar mass, the ratio $f_{\rm tot}$ corresponding to the mass of the most massive star normalized by the total mass, as well as a collision parameter $f_{\rm coll}$ that describes the probability to have collisions within the system. In a visual comparison of the simulation results, we find that the relation and dynamical evolution of these quantities at least is partly related to the initial configuration. Particularly, in simulations that from the start have a significant number of protostars or sink particles already, $f_{*}$ tends to increase when the efficiency $\epsilon_{*}$ increases, as one of the protostars becomes more massive, leading to a higher value of $\epsilon_{*}$ \citep[e.g.,][]{Solar2022, Reinoso2023}. In other simulations that start with the formation of one sink particle, on the other hand, and other sinks then form progressively, $f_{*}$ has an initial value of $1$ but is found to decrease over time while $\epsilon_{*}$ increases as collapse continues and more gas may be converted into stars \citep{Prole2022b, Prole2023, Peters2010, Riaz2020}. Some simulations such as \citet{Chon2020} further show an intermediate behavior, where in part of the simulation $f_{*}$ increases with $\epsilon_{*}$ and in other parts it decreases, as different dynamics may take over at different stages of the evolution.

Considering the relation between $f_{\rm tot}$ and $\epsilon_{*}$, on the other hand, the overall evolution in the simulations appears more similar, and $f_{\rm tot}$ generally is found to increase with $\epsilon_{*}$. The most notable difference here is that this relation is steeper in simulations corresponding to very strong gravitational instability where very massive objects form, as in the cases of \citet{Chon2020}, Solar et al. (2024-in progress) and \citet{Reinoso2023}. We further analyze how $f_{*}$ relates to the collision parameter $f_{\rm coll}$, finding that indeed this parameter is adequate to characterize and distinguish between different regimes, where the simulations of \citet{Chon2020}, \citet{Riaz2020}, Solar et al. (2024-in progress) and \citet{Reinoso2023} correspond to particularly high collision parameters, while \citet{Peters2010} and \citet{Prole2022b, Prole2023} to more moderate or low collision parameters, allowing to estimate how important collisions will be during the evolution of the simulation.

To quantitatively analyze the relation between these dimensionless parameters, we used the Spearman correlation coefficient for assessing the monotonic relationship between the ranks of paired data points. From this analysis, we found a very strong correlation between $f_{\rm tot}$ and $\epsilon_{*}$, a moderately strong correlation between $f_{*}$ and $\epsilon_{*}$. At the same time, $f_{\rm coll}$ has a moderate correlation with $f_{\rm tot}$, a very weak correlation with $f_{*}$ and again a moderate one with $\epsilon_{*}$.

We finally employed different machine learning models that have been trained with the data provided from these simulations, where the best results have been obtained with LightGBM, XGBoost and Random Forest. After training with a subset of the provided simulations, the machine learning models are able to make a good prediction of the parameter $f_{*}$ when $\epsilon_{*}$ and $f_{\rm coll}$ have been provided, suggesting that they provide a good enough indicator for estimating the mass of the most massive object relative to the total stellar mass. We provide various examples where $f_{*}$ is reconstructed as a function of $\epsilon_{*}$ via the trained models, showing a good agreement with the original simulation data. 
%------------------------------------------------------------------------------------------------------------------------------------------------------------%

% 3- Discussion ---------------------------------------------------------------------------------------------------------------------------------------------%
The results obtained here suggest that the outcome of gravitational collapse is not universal, but rather could be divided into different regimes. An approximate characterization of these regimes can be achieved via the collision parameter introduced here, which is roughly defined as the length scale of the system divided by the typical collision length. While in principle in all simulations $f_{\rm tot}$ increases with $\epsilon_{*}$, the steepness of the relation depends specifically on the level of instability characterized through this parameter. We do not claim here that it is the only parameter that affects the formation of massive objects during gravitational collapse, but at least it provides a good description for the data that we analyzed here.

Beyond establishing the relation between $f_{*}$ and $\epsilon_*$, the potential to form very massive objects further depends on the maximum star formation efficiency that can be reached in a given environment \citep{Girichidis2020}. The latter is likely limited by feedback, for example in terms of ionizing radiation \citep[e.g.,][]{Haid2018, Latif2021}, feedback from stellar winds \citep{Das2021} or supernova explosions \citep{Wise2008, Latif2020SN}. The strength of the feedback may even depend on the evolution of the supermassive star itself, particularly if it remains in a bloated state with moderate atmospheric temperatures where radiation feedback remains inefficient for a longer period of time \citep[e.g.,][]{Hosokawa2012, Hosokawa2013, Schleicher2013, Woods2019}, as well as the transition point when the gravitational instability occurs and the star is turned into a massive black hole \citep{Umeda2016, Haemmerle2018}, thereby changing the feedback dynamics. {As in most of the simulations analyzed here, feedback has not been explicitly included, we expect that our results will only hold until feedback significantly affects the dynamics of the collapse.} While the results obtained here do not allow us to determine the maximum star formation efficiency that can be achieved, they at least tell us the expected value of $f_{*}$ for a given value of $\epsilon_{*}$. Even once all gas is expelled as a result of feedback, the latter does not necessarily imply the end of the evolution, but the central massive object may potentially still grow as a result of collisions, as shown in a variety of different studies \citep[e.g.,][]{Sakurai2017, Reinoso2018, Reinoso2020, Vergara2021, Vergara2023}.

{Another limitation concerns the assumption of low turbulent to gravitational energies as well as low to moderate Mach numbers in the initial conditions of the analyzed simulations. This assumption may not always be appropriate, and particularly in cores or clouds with very high turbulent Mach numbers \citep[e.g.][]{McKee2003,Klessen03, Tan2014}, the resulting dynamics could be different. Comparing the dynamics in such environments will certainly be important and interesting, but is also clearly beyond the scope of the current work.
}

%%%% D A T A %%%%%%%%%%%%%%%%%%%%%%%%%%%%%%%%%%%%%%%%%%%%%%%%%%%%%%%%%%%%%%%%%%%%%%%%%%%%%%%%%%%%%%%%%%%%%%%%%%%%%%%%%%%%%%%%%%%%%%%%%%%%%%%%%%%%%%%%%%%%%%%%%%%%      
\section*{Data availability}

The data underlying this article will be shared on reasonable request to the corresponding author.
   
\begin{acknowledgements}
We gratefully acknowledge support by the ANID BASAL project FB21003, as well as via Fondecyt Regular (project code 1201280) and ANID QUIMAL220002. DRGS thanks for funding via the  Alexander von Humboldt - Foundation, Bonn, Germany. RSK acknowledges financial support from the European Research Council via the ERC Synergy Grant ``ECOGAL'' (project ID 855130),  from the German Excellence Strategy via the Heidelberg Cluster of Excellence (EXC 2181 - 390900948) ``STRUCTURES'', and from the German Ministry for Economic Affairs and Climate Action in project ``MAINN'' (funding ID 50OO2206). RSK also thanks for computing resources provided by the Ministry of Science, Research and the Arts (MWK) of the State of Baden-W\"{u}rttemberg through bwHPC and the German Science Foundation (DFG) through grants INST 35/1134-1 FUGG and 35/1597-1 FUGG, and also for data storage at SDS@hd funded through grants INST 35/1314-1 FUGG and INST 35/1503-1 FUGG. KO thanks for funding by the Grants-in-Aid for Basic Research by the Ministry of Education, Science and Culture of Japan (KO:22H00149). BR acknowledges funding through ANID (CONICYT-PFCHA/Doctorado acuerdo bilateral DAAD/62180013), DAAD (funding program number 57451854), and the International Max Planck Research School for Astronomy and Cosmic Physics at the University of Heidelberg (IMPRS-HD). LP acknowledges support from the Irish Research Council Laureate programme under grant number IRCLA/2022/1165, as well as the DiRAC@Durham facility managed by the Institute for Computational Cosmology on behalf of the STFC DiRAC HPC Facility (www.dirac.ac.uk). The equipment was funded by BEIS capital funding via STFC capital grants ST/P002293/1, ST/R002371/1 and ST/S002502/1, Durham University and STFC operations grant ST/R000832/1. DiRAC is part of the National e-Infrastructure. LP also acknowledges the support of the Supercomputing Wales project, which is part-funded by the European Regional Development Fund (ERDF) via Welsh Government. RR thanks for funding through Agencia Nacional de Investigaci\'on y Desarrollo (ANID) (project code SA77210037). PS acknowledges support through ANID/Doctorado en el Extranjero convocatoria 2022 (funding number 72220198), the Federal Ministry of Education and Research and the state governments for supporting this project as part of the joint funding of National High Performance Computing (NHR) and the Kultrun cluster hosted at the Departamento de Astronom\'ia, Universidad de Concepci\'on.
\end{acknowledgements}

%\bibliographystyle{aa}
%\bibliography{astro}

\begin{appendix}
%%%% A P P E N D I X %%% A %%%%%%%%%%%%%%%%%%%%%%%%%%%%%%%%%%%%%%%%%%%%%%%%%%%%%%%%%%%%%%%%%%%%%%%%%%%%%%%%%%%%%%%%%%%%%%%%%%%%%%%%%%%%%%%%%%%%%%%%%%%%%%%%%%%%%%      
\section{ML models \& hyperparameters}\label{AppendixA}
% Figure Appendix-1: Correlation plots for all candidates with heatmap -----------------------------------------------------------------------------------------%
\begin{figure}[!t]
    \centering

    \includegraphics[width=0.78\columnwidth]{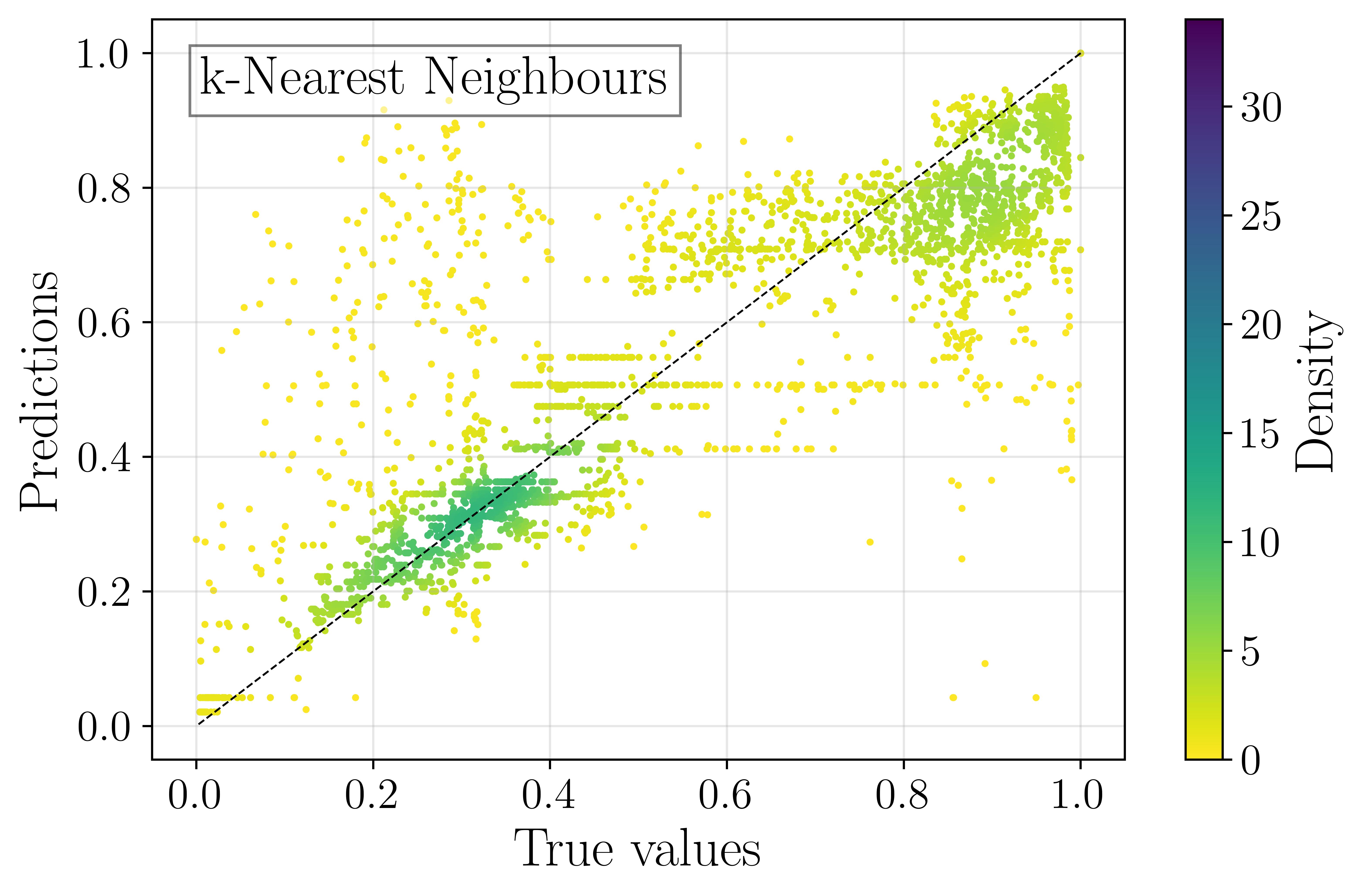}
    \includegraphics[width=0.78\columnwidth]{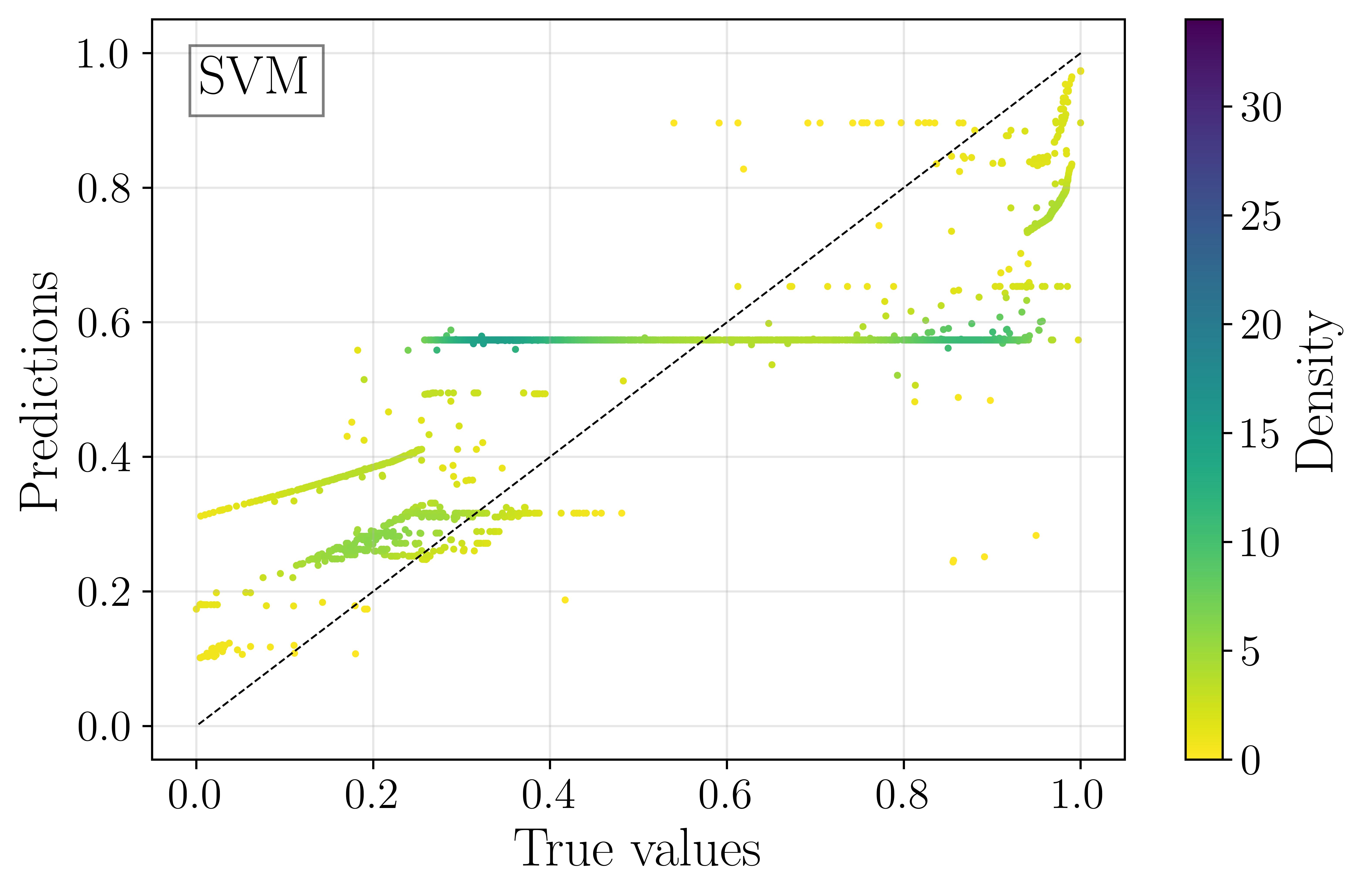}
    \includegraphics[width=0.78\columnwidth]{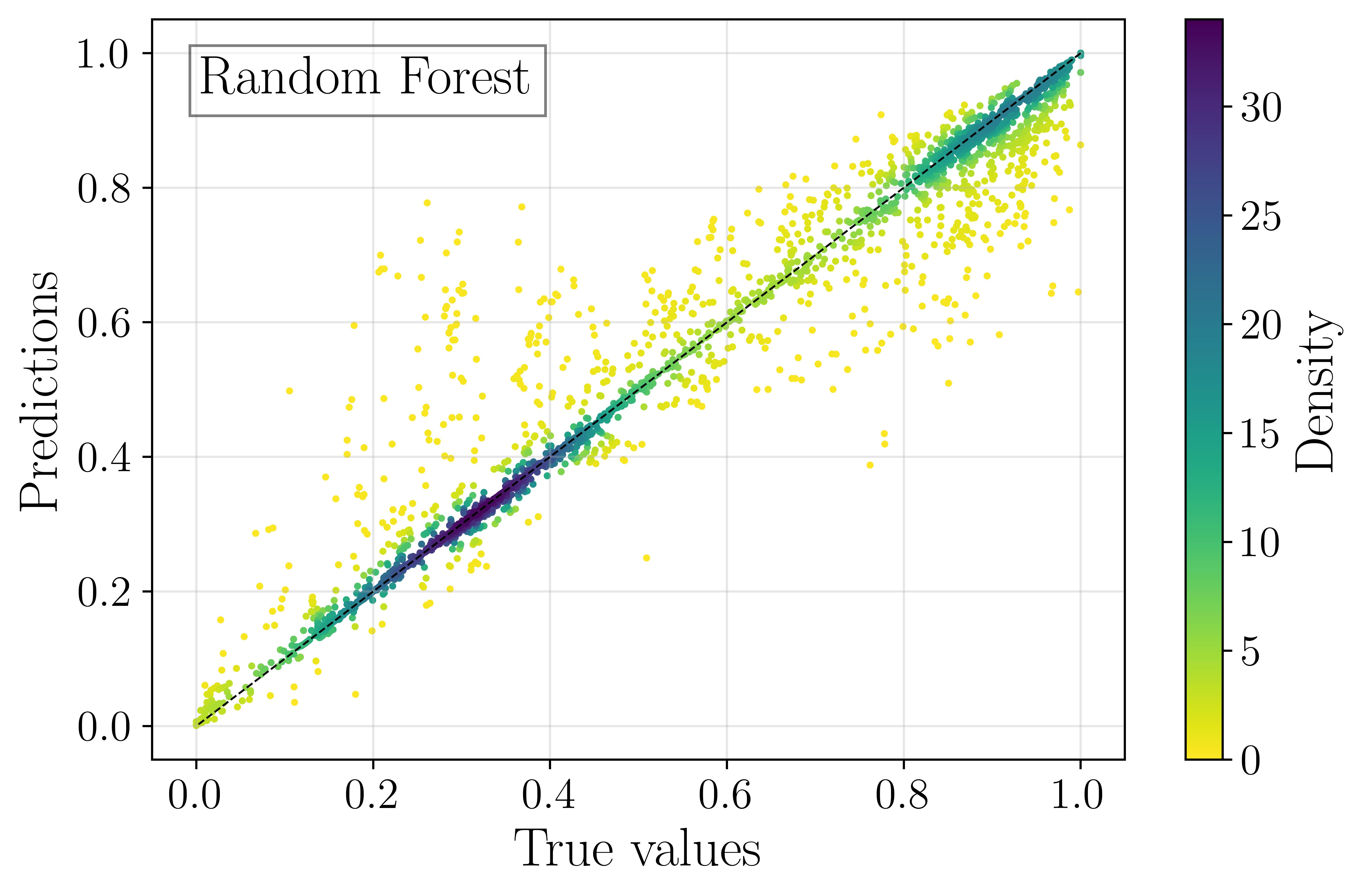}    
    \includegraphics[width=0.78\columnwidth]{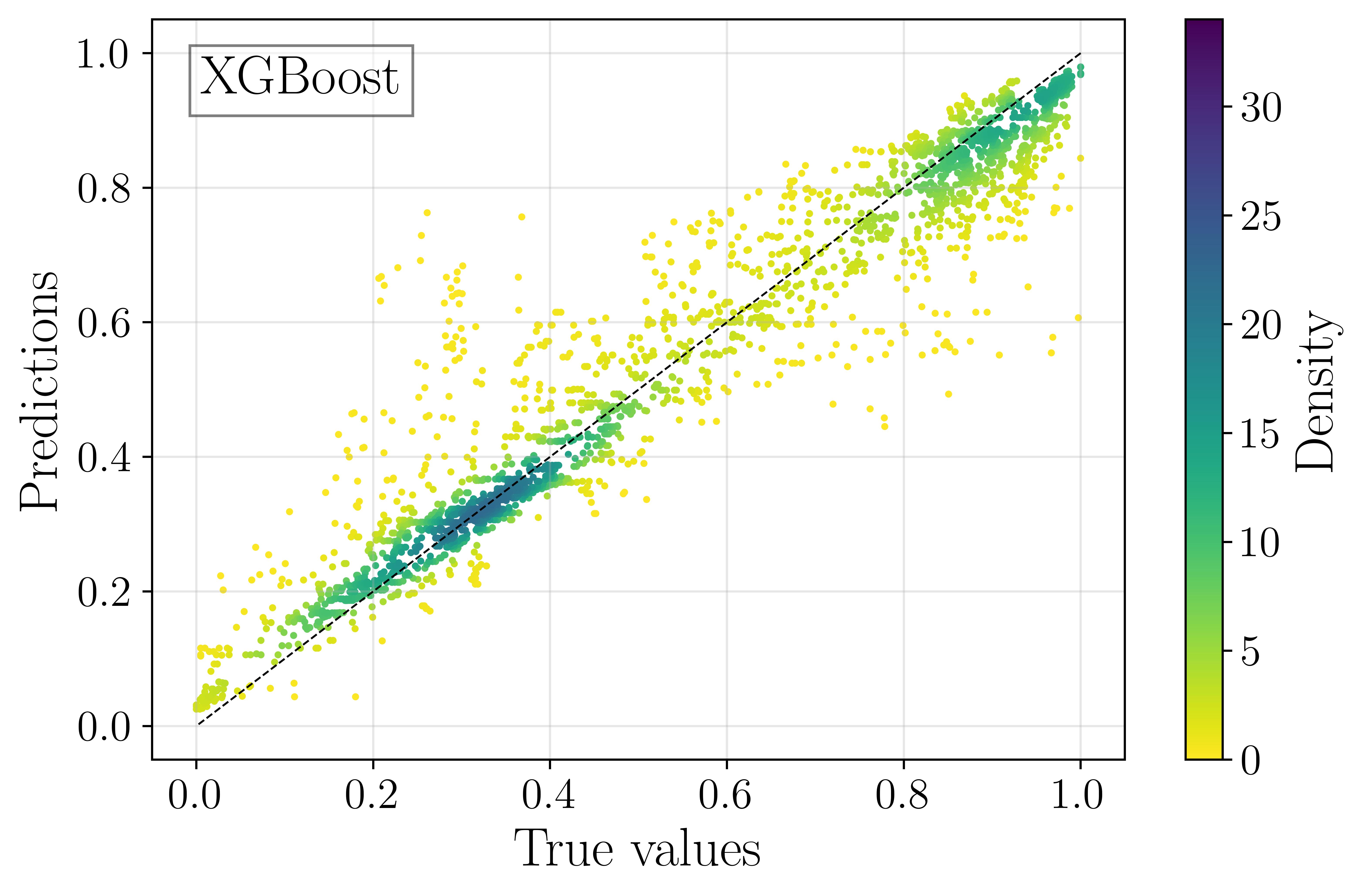}
    \includegraphics[width=0.78\columnwidth]{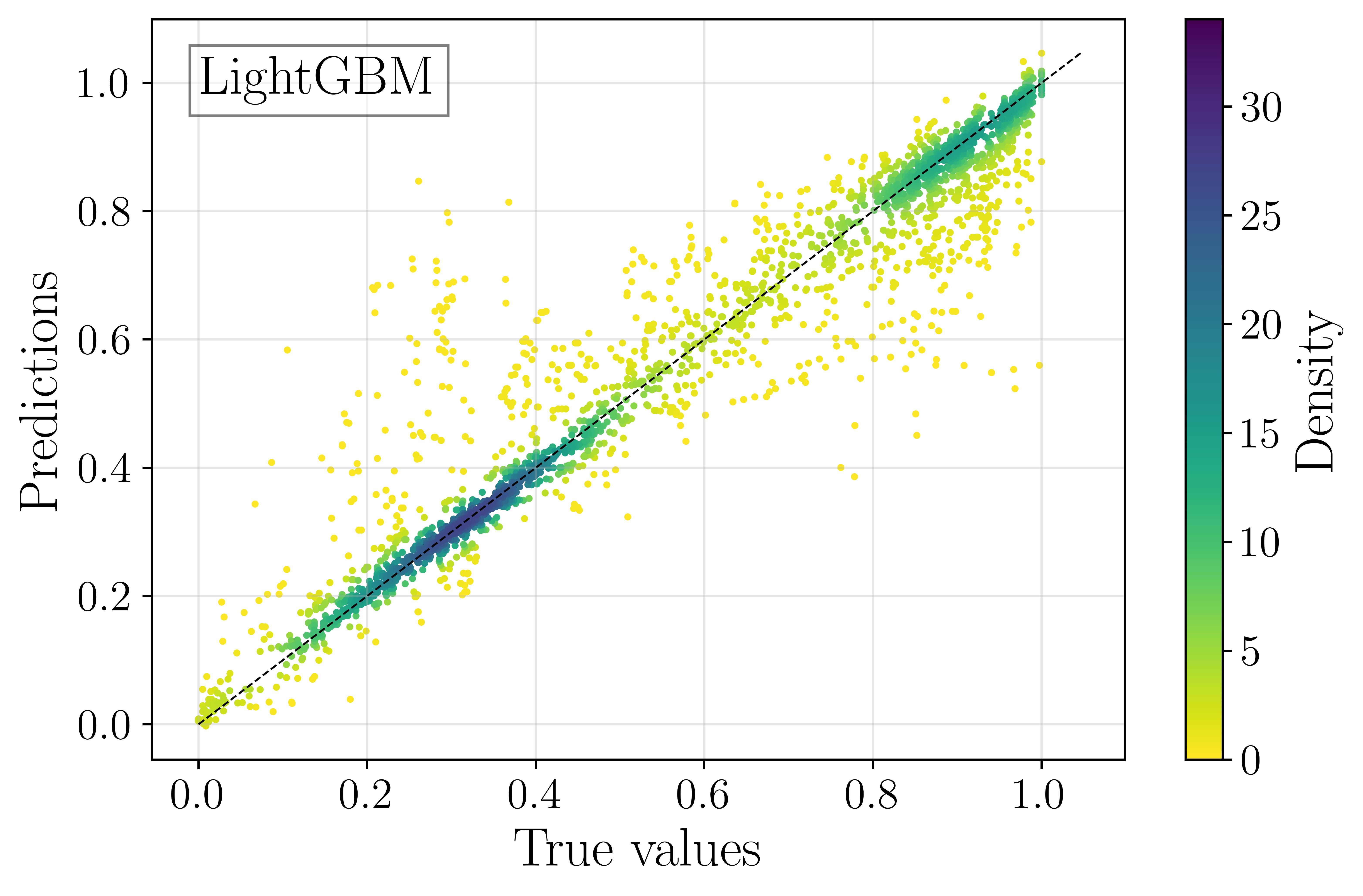}
    \caption{Correlation plots for the test set predictions. Results are shown for all candidates, we also perform a Gaussian kernel density estimation to highlight the regions with more data points.}
    \label{heatmapcorr}
\end{figure}
%---------------------------------------------------------------------------------------------------------------------------------------------------------------%      
In this section we detail each solution obtained at the end of the workflow presented in Sec.~\ref{ML}, including hyperparameters and preprocessing. All solutions were trained from scratch with 3 different splits of data for K-fold cross-validation:
\begin{itemize}
    \item \textbf{k-Nearest Neighbours:}
    \begin{itemize}
        \item Feature extraction: Identity transformation.
        \item Model specifications: $k=30$ and distance defined as the Euclidean distance ($l2$ norm).
        \item Training Score: RMSE$=0.159\pm0.0004$.
        \item Validation Score: RMSE$=0.160\pm0.0033$.
    \end{itemize}
    \item \textbf{SVM:}
    \begin{itemize}
        \item Feature extraction: Identity transformation.
        \item Model specifications: Radial basis function (RBF) kernel with $\gamma=10^{-5}$, and a regularization parameter $C=1000$.
        \item Training Score: RMSE$=0.196\pm0.0022$.
        \item Validation Score: RMSE$=0.212\pm0.0014$.
    \end{itemize}
    \item \textbf{Random Forest:}
    \begin{itemize}
        \item Feature extraction: Standard Scaling of numerical values.
        \item Model specifications: The number of estimator is set to $n=999$ and the minimum number of samples required to be a leaf node is defined as 2.
        \item Training Score: RMSE$=0.054\pm0.0005$.
        \item Validation Score: RMSE$=0.067\pm0.0008$.
    \end{itemize}
    \item \textbf{XGBoost:}
    \begin{itemize}
        \item Feature extraction: $t$-test feature selection, the fraction of total number of features selected from the training dataset is 0.2.
        \item Model specifications: Learning rate $\alpha=0.246$, the minimum loss reduction required to make a further partition $\gamma=0.113$, and a maximum depth for each decision tree of 16.
        \item Training Score: RMSE$=0.064\pm0.0002$.
        \item Validation Score: RMSE$=0.076\pm0.0023$.
    \end{itemize}
    \item \textbf{LightGBM:}
    \begin{itemize}
        \item Feature extraction: $t$-test feature selection, the fraction of total number of features selected from the training dataset is 0.2.
        \item Model specifications: The maximum number of leaves for each decision tree is set to 185, with a maximum depth of 180.
        \item Training Score: RMSE$=0.067\pm0.0004$.
        \item Validation Score: RMSE$=0.076\pm0.0056$.
    \end{itemize}
\end{itemize}

Finally, in Fig.~\ref{heatmapcorr}, we provide the correlation plots for the predictions of $f_*(t)$ for all models. The summary of the metrics relative to these predictions with the test set can be found in Table~\ref{tabmetrics}. 

All the solutions are programmed in Python 3 (version: 3.12.2) under the framework of AutoML, which is executed automatically. However, the implementations of the models come from open-source Python packages, including scikit-learn\footnote{Scikit-Learn: https://scikit-learn.org/stable/index.html}, Faiss\footnote{Faiss: https://faiss.ai/}, LightGBM\footnote{LightGBM: https://lightgbm.readthedocs.io/en/}, and XGBoost\footnote{XGBoost: https://xgboost.readthedocs.io/en/}. The trained models and respective statistical weights shown in this article will be shared upon reasonable request.
\end{appendix}


%\begin{thebibliography}{}

%\end{thebibliography}

\end{document}